# Direct observation of 3D atomic packing in monatomic amorphous materials

Yakun Yuan[1*], Dennis S. Kim[1*], Jihan Zhou[1*], Dillan J. Chang[1], Fan Zhu[1], Yasutaka Nagaoka[2], Yao Yang[1], Minh Pham[3], Stanley J. Osher[3], Ou Chen[2], Peter Ercius[4], Andreas K. Schmid[4], Jianwei Miao[1†]

*1Department of Physics & Astronomy and California NanoSystems Institute, University of California, Los Angeles, CA 90095, USA. 2Department of Chemistry, Brown University, Providence, RI 02912, USA. 3Department of Mathematics, University of California, Los Angeles, CA 90095, USA. 4National Center for Electron Microscopy, Molecular Foundry, Lawrence Berkeley National Laboratory, Berkeley, CA 94720, USA.*

*\*These authors contributed equally to this work.*

*†Email: miao@physics.ucla.edu*

**Liquids and solids are two fundamental states of matter. However, due to the lack of direct experimental determination, our understanding of the 3D atomic structure of liquids and amorphous solids remained speculative. Here we advance atomic electron tomography to determine for the first time the 3D atomic positions in monatomic amorphous materials, including a Ta thin film and two Pd nanoparticles. We observe that pentagonal bipyramids are the most abundant atomic motifs in these amorphous materials. Instead of forming icosahedra, the majority of pentagonal bipyramids arrange into networks that extend to medium-range scale. Molecular dynamic simulations further reveal that pentagonal bipyramid networks are prevalent in monatomic amorphous liquids, which rapidly grow in size and form icosahedra during the quench from the liquid state to glass state. The experimental**

**method and results are expected to advance the study of the amorphous-crystalline**

**phase transition and glass transition at the single-atom level.**

In 1952, Frank hypothesized that icosahedral order is the prevalent atomic motif in monatomic liquids[1]. Over the past six decades, there have been a great deal of experimental, computational, and theoretical studies to understand the structure of liquids and amorphous materials[2-21]. A polytetrahedral packing model was proposed to explain the 3D atomic structure of monatomic liquids and amorphous materials[22], in which icosahedral order is a key feature. The icosahedral order has also been found to play a critical role in the structure of metallic glasses and quasicrystals[23-29]. Despite all these developments, however, no experimental method could directly determine the 3D atomic packing of liquids and amorphous materials due to the lack of long-range order. Atomic electron tomography (AET), allowing the determination of 3D atomic structure of materials without assuming crystallinity[30], is uniquely positioned to address this challenge. Since its first demonstration in 2012[31], AET has been applied to reveal a wide range of crystal defects in materials such as grain boundaries, dislocations, stacking faults, point defects, atomic ripples, bond distortion, strain tensors and chemical order/disorder with high 3D precision[30,32-39]. More recently, AET was used to determine the structure of a multi-component metallic glass nanoparticle and quantitatively characterize the short- and medium-range order of its 3D atomic arrangement[40]. Here we advance AET to reveal the 3D atomic structure of an amorphous Ta thin film and two amorphous Pd nanoparticles that are not metallic glasses but liquid-like solids. We observe that pentagonal bipyramids are the main atomic motifs in the monatomic amorphous materials. Instead of assembling icosahedra, most pentagonal bipyramids closely connect with each other to form a novel medium-range order named the pentagonal bipyramid network (PBN).

**Atomic electron tomography of monatomic amorphous materials**

The AET experiments were conducted with a scanning transmission electron microscope in annular dark-field mode. Tomographic tilt series were acquired from an amorphous Ta thin film and two Pd nanoparticles (Supplementary Table 1 and Supplementary Figs. 1-3), which were synthesized by physical vapor deposition and colloidal chemistry with ligand engineering, respectively (Methods). Images acquired before, during and after the acquisition of each tilt series indicate a minimal change of the sample structure throughout the experiment (Supplementary Fig. 4). To verify the amorphous nature of the samples, 2D power spectra were calculated from the experimental images, showing the amorphous halo (insets in Supplementary Figs. 1-3). Electron diffraction experiments were also conducted to obtain the structure factors and pair distribution functions (PDFs) of the Ta film and Pd nanoparticles[41], further confirming the amorphous nature of the samples (Methods, Supplementary Fig. 5). After image pre-processing, each tilt series was reconstructed by a real space iterative algorithm and the 3D coordinates of individual atoms were traced and refined to produce an experimental atomic model (Methods). Compared to nanoparticles[30,36,37], the AET reconstruction of thin films is more challenging as the projections at different tilt angles contain different volumes of the thin film (Supplementary Fig. 1). We have developed a real space iterative reconstruction algorithm to solve this problem and determine the 3D atomic coordinates in the amorphous Ta thin film (Methods). The precision of the 3D atomic coordinates was validated to be 18 picometers (Supplementary Fig. 6).

Figure 1a, b and Supplementary Fig. 7 show the experimental 3D atomic model of the Ta thin film and two Pd nanoparticles (named $Pd_1$ and $Pd_2$), respectively. To quantify the disorder, we calculated the bond orientational order parameters for all the atoms[42] (Methods, Supplementary Fig. 8a-c). We find that 20.1%, 2.2% and 1.8% of the atoms form crystal nuclei on the surface of the Ta thin film, $Pd_1$ and $Pd_2$ nanoparticles, respectively (grey atoms in Fig. 1a, b, Supplementary Fig. 7a, and Supplementary Videos 1-3). After excluding these nuclei, we plot the PDFs of the disordered atoms (Fig. 1c), which exhibit similar shapes despite different chemical composition and synthesis methods of the samples. As a comparison, the PDF of a Ta liquid obtained by the molecular dynamics (MD) simulation is shown as a dotted curve in Fig. 1c, in which the peak and valley positions agree with those of the Ta thin film and two Pd nanoparticles.

Next, we quantified the 3D short-range atomic packing of the samples using the Voronoi tessellation[4,23]. This method characterizes each local polyhedron around a centre atom by calculating a Voronoi index, $<n_3, n_4, n_5, n_6>$, where $n_i$ denotes the number of $i$-edged faces (Methods). Figure 1d shows the 12 most populated Voronoi polyhedra in the three samples. We find that distorted icosahedra with $<0,0,12,0>$, $<0,1,10,2>$, $<0,2,8,2>$ and $<0,2,8,1>$ account for 9.8% in the three samples. In contrast, among all the faces in the Voronoi polyhedra (Supplementary Fig. 8d), five-edged faces are the most abundant (45.5%), indicating that the majority of five-edged faces do not form distorted icosahedra. From the Voronoi indices, we determined the average coordination number of the Ta, $Pd_1$ and $Pd_2$ sample to be 12.2, 12.3 and 12.3, respectively, which agree well with that of monatomic liquids (12±1) measured by diffraction experiments[43].

**Polytetrahedral packing in monatomic amorphous materials**

To quantitatively characterize the tetrahedra in the amorphous materials (Methods), we used the distortion parameter[14,44], defined as $\delta = e_{max}/e_{avg} - 1$, where $e_{max}$ and $e_{avg}$ are the maximum and average edge length of each tetrahedron, respectively. Supplementary Fig. 9 (green curves) shows the fraction of the tetrahedra as a function of $\delta$. With $\delta > 0.2$, more than 96.8% of the atoms in the samples form tetrahedra. By sharing faces, these tetrahedra constitute four main motifs: triplets, quadrilateral, pentagonal and hexagonal bipyramids (Fig. 2a). The four motifs are represented by three-, four-, five- and six-fold skeletons, which are formed by connecting the centroids of the tetrahedra (coloured lines in Fig. 2a). Figure 2b and Supplementary Fig. 9 show the fraction of the four motifs as a function of $\delta$. With $\delta < 0.2$, the tetrahedra are not fully packed in 3D space leading to a dominant fraction of triplets. With larger $\delta$, the populations of quadrilateral, pentagonal and hexagonal bipyramids increase, while triplets decrease. In the following analysis, we choose $\delta \leq 0.255$, which was previously used in mathematical and numerical simulation studies[14,44]. Figure 2c shows the population of the four motifs in the three samples, indicating that pentagonal bipyramids are the most abundant atomic motifs. This observation is consistent with the Voronoi tessellation analysis (Supplementary Fig. 8d), and can be explained by the fact that the atomic packing in pentagonal bipyramids requires less distortion than in the other motifs[22,24].

Since a tetrahedron and a pentagonal bipyramid represent the densest packing of four and seven atoms[22], respectively, we correlated polytetrahedral packing with the local mass density of the amorphous materials (Methods). Figure 3a, Supplementary Fig. 10a and d show the mass density distribution in the regions of three samples with and without polytetrahedral packing, where the average mass density increases with polytetrahedral packing. We also observe 3D local mass density heterogeneity in the amorphous materials. A slice through each sample shows the local mass density heterogeneity overlaid with polytetrahedral packing (Fig. 3b, Supplementary Fig. 10b and e). A magnified region in each sample reveals that 3D local mass density heterogeneity is strongly correlated to the atomic packing of the four motifs in the three samples (Fig. 3c, Supplementary Fig. 10c and f).

As pentagonal bipyramids are the most abundant motifs, we quantified their 3D atomic packing in the three amorphous materials. Supplementary Fig. 11a-c shows the bond angle distribution of the Ta film, $Pd_1$ and $Pd_2$ nanoparticles, respectively, which agrees with the previous study of liquid metals using reverse Monte Carlo simulations[12]. The two peaks of the bond angle distribution are consistent with the internal angle of a tetrahedron and a pentagon, respectively (Supplementary Fig. 11d). Next, we quantified the three bonds in the pentagonal bipyramid: the capping atom bond ($\alpha$), the capping-ring atom bond ($\beta$) and the ring atom bond ($\gamma$) (Fig. 4a, top). According to the polytetrahedral packing model[22,24], an ideal pentagonal bipyramid consisting of seven atoms has the $\alpha$ bond 5% longer than the $\beta$ and $\gamma$ bonds. However, we find that the $\alpha$ bond is statistically 2.5% longer than the $\gamma$ bond and the $\beta$ bond is 1.3% shorter than the $\gamma$ bond in the three amorphous samples (Fig. 4b). We also observe that the angle ($\theta$) between the $\alpha$ bond and the plane of five ring atoms (Fig. 4a, bottom) deviates from an ideal pentagonal bipyramid of $\theta = 0°$. Figure 4c shows that the average $\theta$ was measured to be $10.0°$, $10.7°$ and $10.9°$ for the amorphous Ta film, $Pd_1$ and $Pd_2$ nanoparticles, respectively. All these results indicate that the pentagonal bipyramids are distorted in these amorphous samples. We also observe the same distortion in an MD simulated Ta liquid (Methods). As the liquid is quenched from 5200 K to 300 K, $\theta$ decreases (Fig. 4d) but the $\alpha/\gamma$ and $\beta/\gamma$ ratios remain relatively unchanged (Fig. 4b).

**The pentagonal bipyramid network**

We find that a number of pentagonal bipyramids link to each other by sharing four or five atoms with their neighbours (Fig. 5a and b), which we define as vertex- or edge-sharing of the five-fold skeletons, respectively. Figure 5c and Supplementary Fig. 12 show the fraction of pentagonal bipyramids as a function of the number of vertex- and edge-sharing neighbours. We observe in the three samples that 63.5% of bipyramids do not share any vertex with their neighbours, but the majority of them (72.5%) have at least one edge-sharing neighbour. Figure 5d and e shows two pentagonal bipyramid clusters with the most vertex- and edge-sharing neighbours, respectively, where the larger cluster is formed by edge-sharing. These results indicate that edge-sharing of the five-fold skeletons is a more dominant feature in the packing of pentagonal bipyramids. We then investigate if these pentagonal bipyramids form icosahedra. An icosahedron requires the packing of 12 pentagonal bipyramids by edge-sharing. Due to geometric frustration[22,27], the five-fold skeletons in an icosahedron form a regular dodecahedron with the dihedral angle ($\varphi$) of 116.57° (Fig. 5f, inset). But we observe that the dihedral angles ($\varphi$) between two edge-sharing pentagonal bipyramids peak at 120.7° in the three amorphous samples (Fig. 5f), which is close to $\varphi = 120°$ in the absence of geometric frustration. This observation further confirms that the pentagonal bipyramids only assemble partial icosahedra (Fig. 5g). These results do not contradict that 9.8% of all the Voronoi polyhedra in the three samples are distorted icosahedra because the vast majority of these distorted icosahedra have a large distortion with $\delta > 0.255$ (Supplementary Fig. 13). When choosing $\delta \leq 0.255$, the total numbers of distorted icosahedra and pentagonal bipyramids in the three amorphous materials are 17 and 26262, respectively, showing that the pentagonal bipyramids are far more abundant than the distorted icosahedra in the samples.

Instead of assembling icosahedra, most pentagonal bipyramids with edge-sharing skeletons form PBNs in these amorphous samples. Figure 5h shows a representative PBN, which consists of five partial icosahedra. Figure 5i, Supplementary Figs. 14 and 15 show the histograms of the PBNs as a function their size and length. The largest PBN is found in the Ta thin film, which consists of 135 pentagonal bipyramids formed by 165 atoms with an end-to-end length of 2.83 nm (Fig. 5j). The five largest PBNs in the Ta, $Pd_1$ and $Pd_2$ samples are shown in Supplementary Figs. 16-18. Compared with the networks formed by quadrilateral and hexagonal bipyramids, the PBNs not only are more abundant, but also have a larger size in all three samples (Supplementary Figs. 14 and 15), indicating the PBNs are dominant in monatomic amorphous materials.

To investigate if PBNs are prevalent in other amorphous systems such as liquids and metallic glasses, we employed MD simulations using the large-scale atomic/molecular massively parallel simulator[45] (Methods). A bulk Ta solid was melted at 5200 K, quenched at a cooling rate of $10^{13}$ K/s and brought to equilibrium at 300 K. The PDFs of the Ta structures at varying temperatures are shown in Supplementary Fig. 19. At 5200 K, the PDF of the Ta liquid resembles those of experimentally measured amorphous materials in terms of the peak and valley positions (dotted curve in Fig. 1c). At 300 K, the splitting of the $2^{nd}$ and $3^{rd}$ peaks in the PDF indicates the formation of the Ta metallic glass[18,24] (arrows in Supplementary Fig. 19). By analysing the polytetrahedral packing of these Ta structures with $\delta \leq 0.255$, we find that pentagonal bipyramids are the most abundant atomic motifs across the entire temperature range and their population dramatically increases with the decrease of the temperature (Fig. 6a). At 5200 K, we observe PBNs and partial icosahedra in the Ta liquid (Fig. 6b, c and Supplementary Fig. 20). These PBNs extend to medium-range scale and resemble those found in the experimental Ta, $Pd_1$ and $Pd_2$ samples (Fig. 5i, j and Supplementary Figs. 16-18). With the decrease of temperature, the pentagonal bipyramids form a fraction of icosahedra and the PBNs rapidly grow in size (Fig. 6b, d and Supplementary Fig. 21). At 300 K, a huge PBN is created across the entire Ta metallic glass (Supplementary Fig. 21c) and more icosahedra are formed (Fig. 6b and e). We have also performed MD simulations of quenching Pd from a liquid to a metallic glass state and observed very similar results.

**Discussion and outlook**

Our experimental results, coupled with MD simulations, provide a fundamental insight into the 3D atomic packing of monatomic amorphous materials and liquids. Although there are several crystal nuclei on the surface of the Ta thin film and two Pd nanoparticles, we verify that the crystal nuclei have a minimal impact on the structural disorder of the rest of the samples based on the following three observations. First, after removing the crystal nuclei, the peak and valley positions of the PDFs of the samples are in good agreement with those of the MD simulated Ta liquid (Fig. 1c). Second, after quantitatively analysing the crystalline-amorphous interface in the samples, we find that the characteristic width of the interface is around 3.0-4.3 Å (Methods, Supplementary Fig. 22), which is consistent with previous MD simulation results[46]. This analysis indicates that the crystal nuclei do not affect the atomic-level structural disorder beyond a few angstroms. Third, the 3D atomic packing in the amorphous regions of the Ta thin film and Pd nanoparticles resembles that of an MD simulated Ta liquid (Supplementary Figs. 16-18 and 20), exhibiting features completely different from the crystalline structure. This observation also poses a deep question on why these monatomic amorphous materials and liquid, representing two different states of matter, have similar 3D atomic structures.

Our experimental and MD simulation results further reveal that pentagonal bipyramids are the prevalent atomic motifs and form a medium-range order (PBNs) in monatomic amorphous materials and liquids. During the quench from a liquid to metallic glass state, the PBNs quickly grow in size and assemble a fraction of icosahedra, indicating an important role of the PBNs during the glass transition. Looking forward, the ability to determine the 3D atomic structure of amorphous thin films is expected to greatly expand the applicability of AET to a broad range of technologically relevant materials[47]. Moreover, the experimental method and results reported in this work could have an important impact on different fields, ranging from the direct determination of the 3D atomic structure of quasicrystals[28,29] to the study of the physics of jamming[48], the amorphous-crystalline phase transition and glass transition at the atomic scale[49,50].

**Acknowledgements** We thank Jim Ciston for help with data acquisition and Xuezeng Tian with data analysis. Funding: This work was primarily supported by the US Department of Energy (DOE), Office of Science, Basic Energy Sciences, Division of Materials Sciences and Engineering under award DE-SC0010378. We also thank the support by STROBE: a National Science Foundation (NSF) Science and Technology Center under award DMR-1548924. Some of the data analysis was partially supported by the NSF DMREF program under award DMR-1437263, and the Army Research Office MURI program under grant no. W911NF-18-1-0431. The ADF-STEM imaging with TEAM I was performed at the Molecular Foundry, which is supported by the Office of Science, Office of Basic Energy Sciences of the US DOE under contract no. DE-AC02-05CH11231.

## Figures and Figure legends

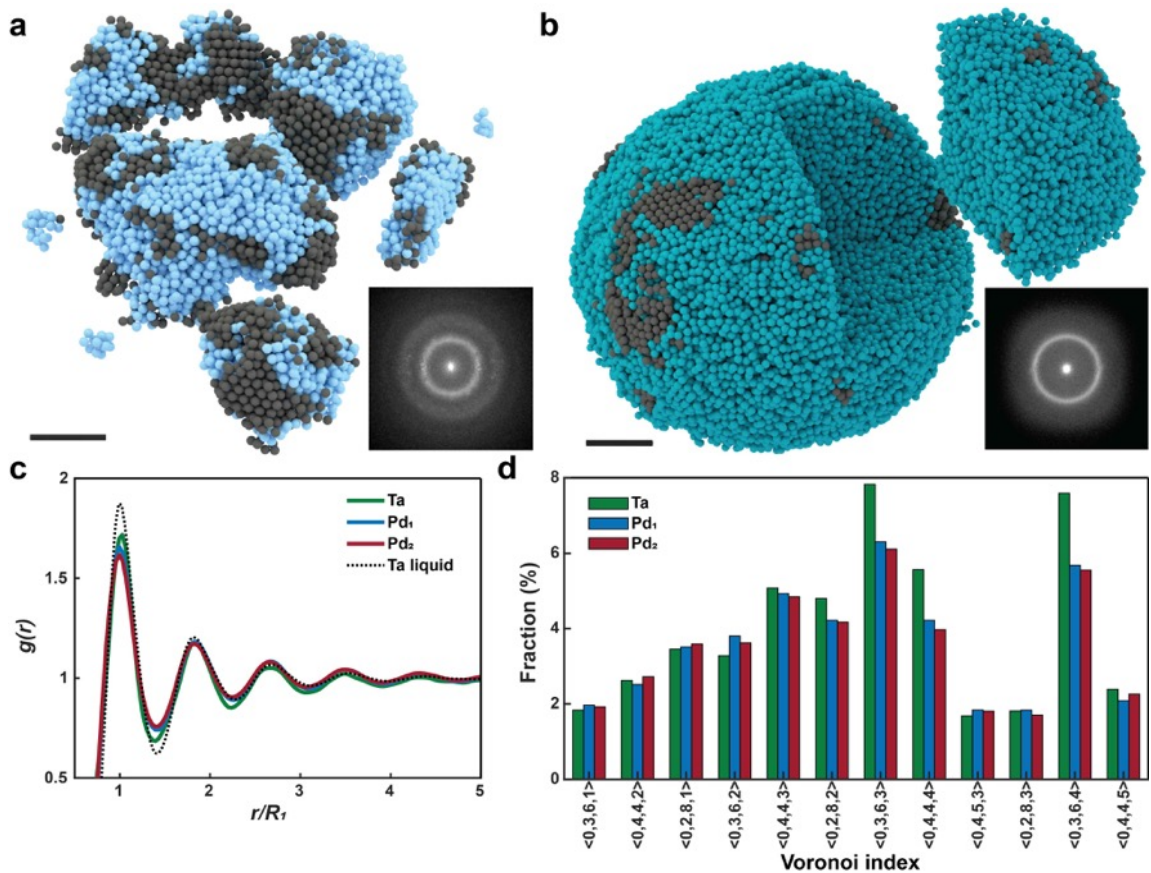

**Fig. 1 | Determination of the 3D atomic structure of monatomic amorphous materials.** Experimental 3D atomic model of an amorphous Ta film (**a**) and a Pd nanoparticle (Pd$_1$) (**b**) with surface crystal nuclei in grey. Scale bar, 2 nm. The two insets show the average 2D power spectra of the experimental images for the Ta film and Pd$_1$ nanoparticle, where the amorphous halo is visible. **c**, PDFs of the Ta film (green), two Pd nanoparticles (Pd$_1$ in blue and Pd$_2$ in red) and an MD simulated Ta liquid at 5200 K (dotted curve). The PDFs were normalized by the Ta and Pd atom size. **d**, 12 most populated Voronoi polyhedra in the three samples, where the Voronoi index is arranged with the increase of the coordination number.

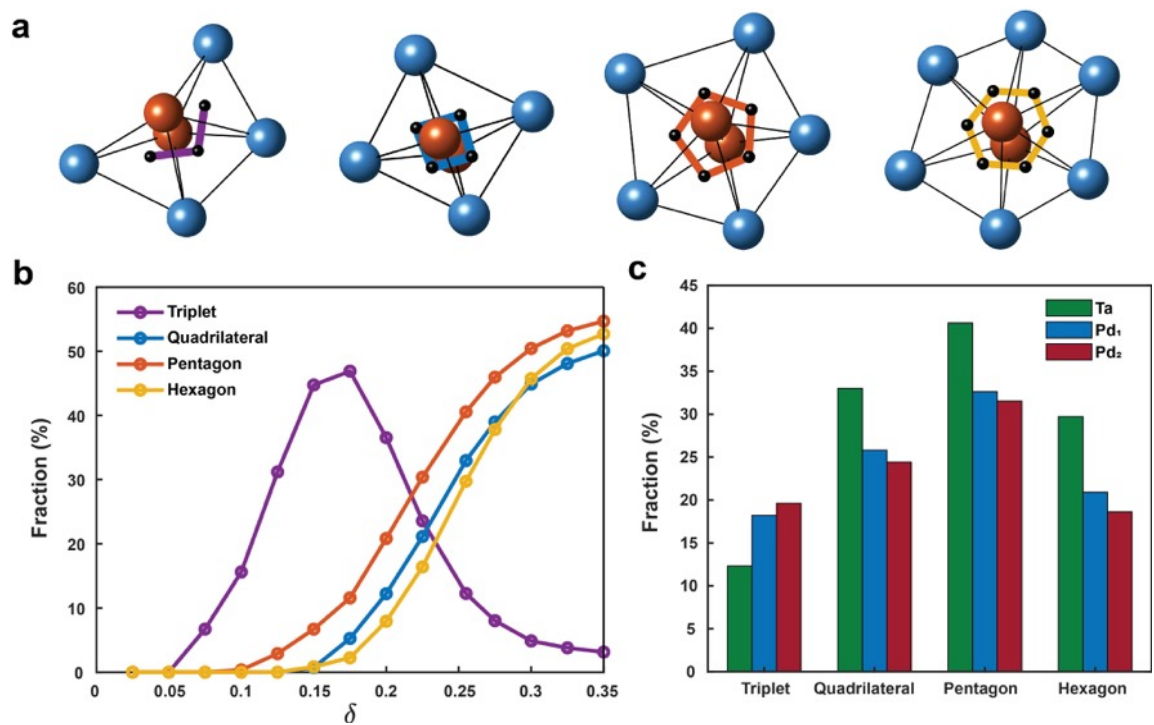

**Fig. 2 | Polytetrahedral packing in the amorphous Ta film and Pd nanoparticles. a**, Four most populated atomic motifs (triplets, quadrilateral, pentagonal and hexagonal bipyramids) in the three samples, where the capping atoms are in brown and the ring atoms are in blue, connected by the bonds (black lines). The four motifs are represented by a three- (purple), four- (blue), five- (orange) and six-fold skeleton (yellow), which

connect the centroid (black dot) of each tetrahedron. **b**, Population of the four atomic motifs in the Ta film as a function of $\delta$. **c**, Distribution of the four atomic motifs in the amorphous Ta film (green), $Pd_1$ (blue) and $Pd_2$ (red) nanoparticles with $\delta \leq 0.255$.

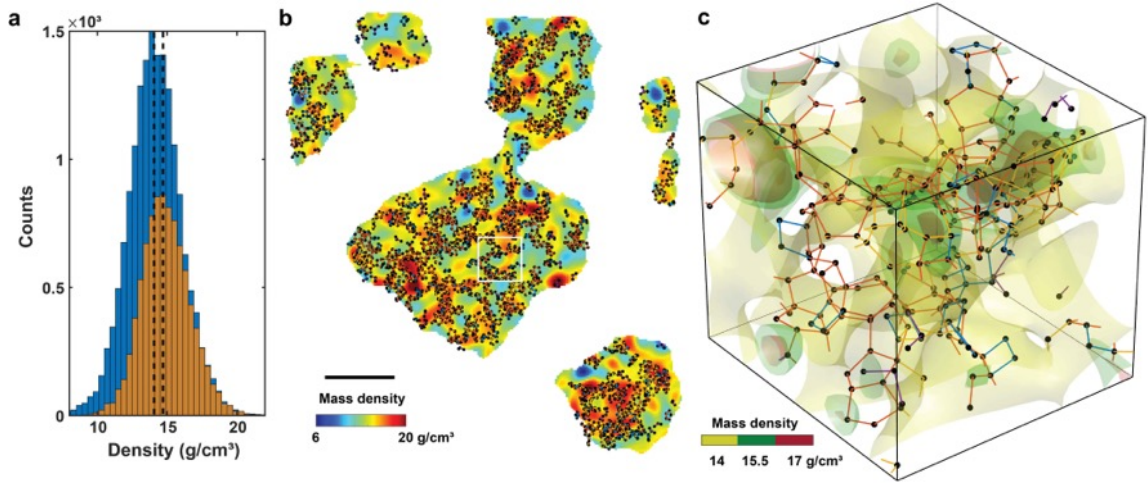

**Fig. 3 | Correlation of 3D local mass density heterogeneity and polytetrahedral packing. a**, Mass density distribution in the regions of the amorphous Ta film with (yellow) and without polytetrahedral packing (blue), where polytetrahedral packing increases the average mass density by 3% (dashed lines). **b**, A slice through the Ta film shows the local mass density heterogeneity (colour) overlaid with polytetrahedral packing (black). **c**, 3D surface rendering of local mass density heterogeneity magnified from the square region in (**b**), which is overlaid with three- (purple), four- (blue), five- (orange) and six-fold (yellow) skeletons. Scale bar, 2 nm.

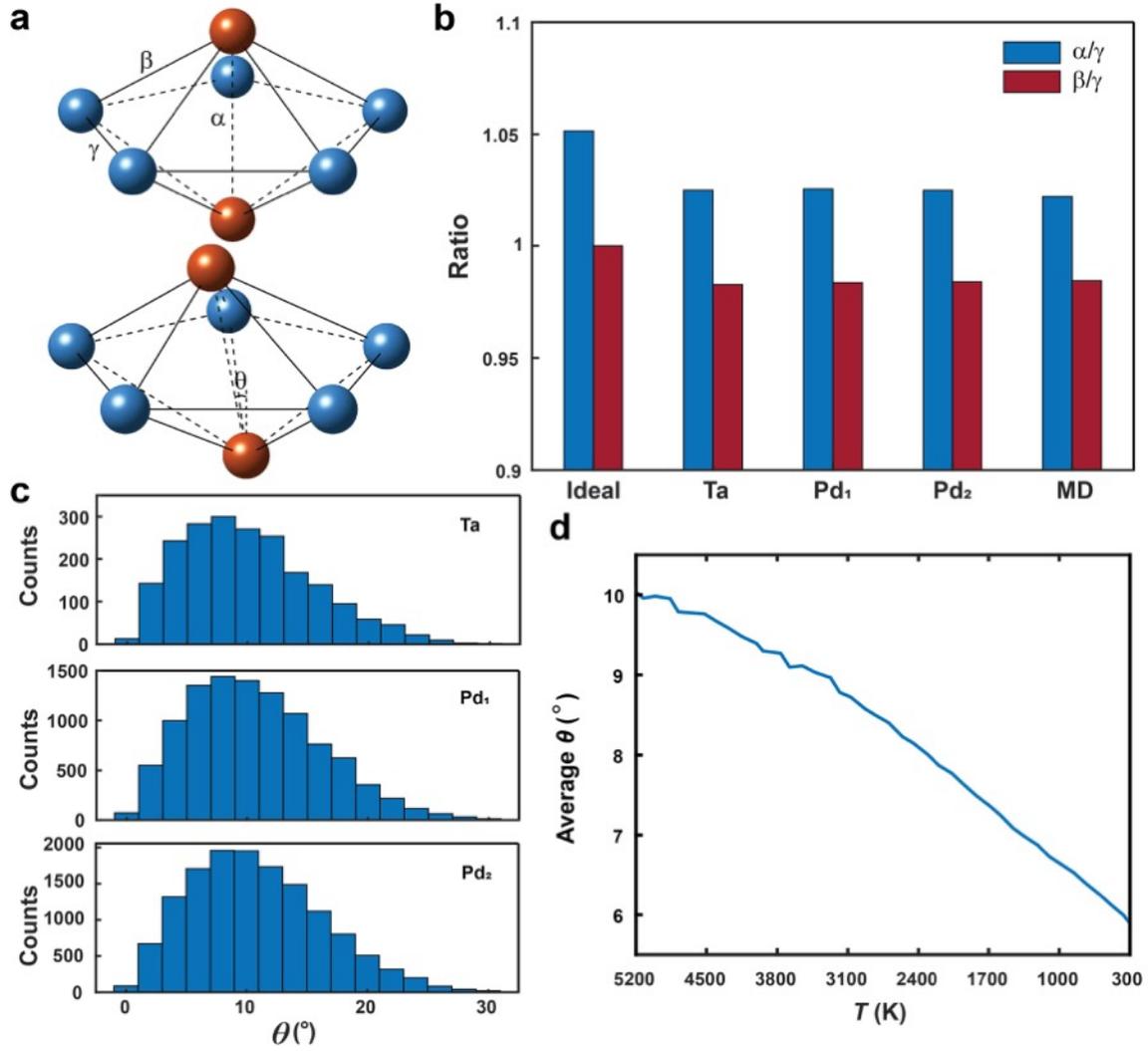

**Fig. 4 | Quantitative characterization of 3D atomic packing of pentagonal bipyramids. a**, An ideal pentagonal bipyramid (top) consisting of two capping (brown) and five ring atoms (blue). α, β and γ represent the capping, capping-ring and ring atom bonds, respectively. Bottom shows the average pentagonal bipyramid of the amorphous Ta film, where the $\alpha$ bond and the plane of five ring atoms form an angle ($\theta$). **b**, The $\alpha/\gamma$ and β/γ ratios of an ideal pentagonal bipyramid as well as those in the amorphous Ta film, $Pd_1$, $Pd_2$ nanoparticles and MD simulated Ta liquid at 5200 K. **c**, Distribution of $\theta$ in the Ta film (top), $Pd_1$ (middle) and $Pd_2$ (bottom) nanoparticles with the average $\theta$ of 10.0°,

10.7° and 10.9°, respectively. **d**, Average $\theta$ as a function of the temperature during the quench of Ta from the liquid to metallic glass state.

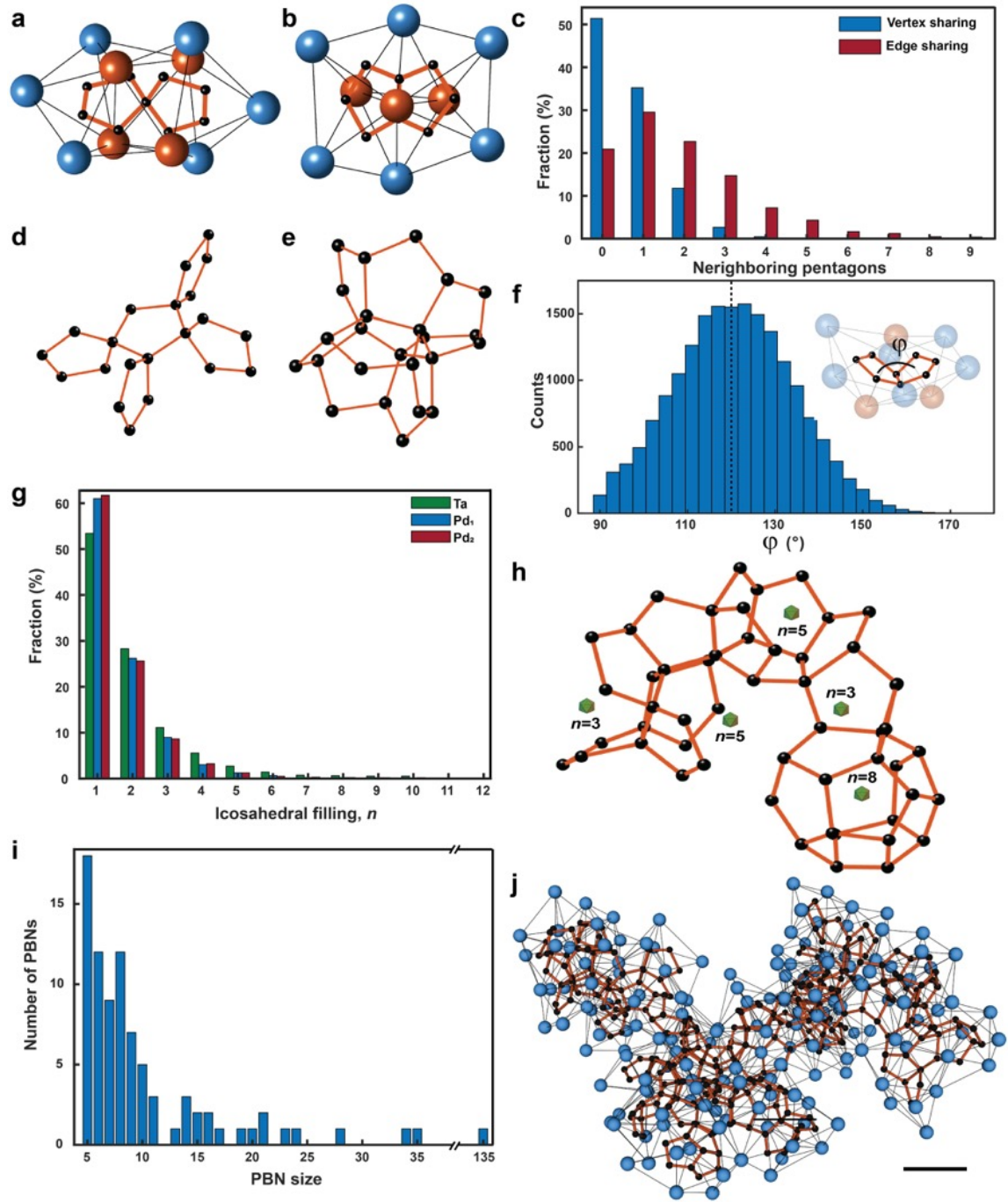

**Fig. 5 | Direct observation of pentagonal bipyramid networks in monatomic amorphous materials. a**, **b**, Vertex- and edge-sharing of the five-fold skeletons, which share four and five atoms with their neighbours, respectively. The capping atoms are in

brown. **c**, Population of pentagonal bipyramids as a function of the number of vertex- and edge-sharing neighbours in the amorphous Ta film. Pentagonal bipyramid clusters with the most vertex- (**d**) and edge-sharing neighbours (**e**) in the three amorphous materials. **f**, Distribution of the dihedral angles ($\varphi$) between two edge-sharing pentagonal bipyramids (inset) in the three amorphous materials, where the average $\varphi$ is 120.7° (dashed line). **g**, Population of pentagonal bipyramids filling partial icosahedral sites ($n$). The formation of a full icosahedron requires $n = 12$. **h**, Five-fold skeleton of a representative PBN, containing 5 partial icosahedra with n = 3, 5, 5, 3 and 8. The centre of each partial icosahedron is labelled by an icosahedron (green). **i**, Histogram of the PBNs as a function of their size (i.e. the number of pentagonal bipyramids) in the Ta thin film. **j**, The largest PBN in the amorphous materials, containing 135 pentagonal bipyramids formed by 165 Ta atoms (blue balls). The black lines represent the bonds between the Ta atoms and the black dots and orange lines show the PBN with five-fold skeletons. Scale bar, 4 Å.

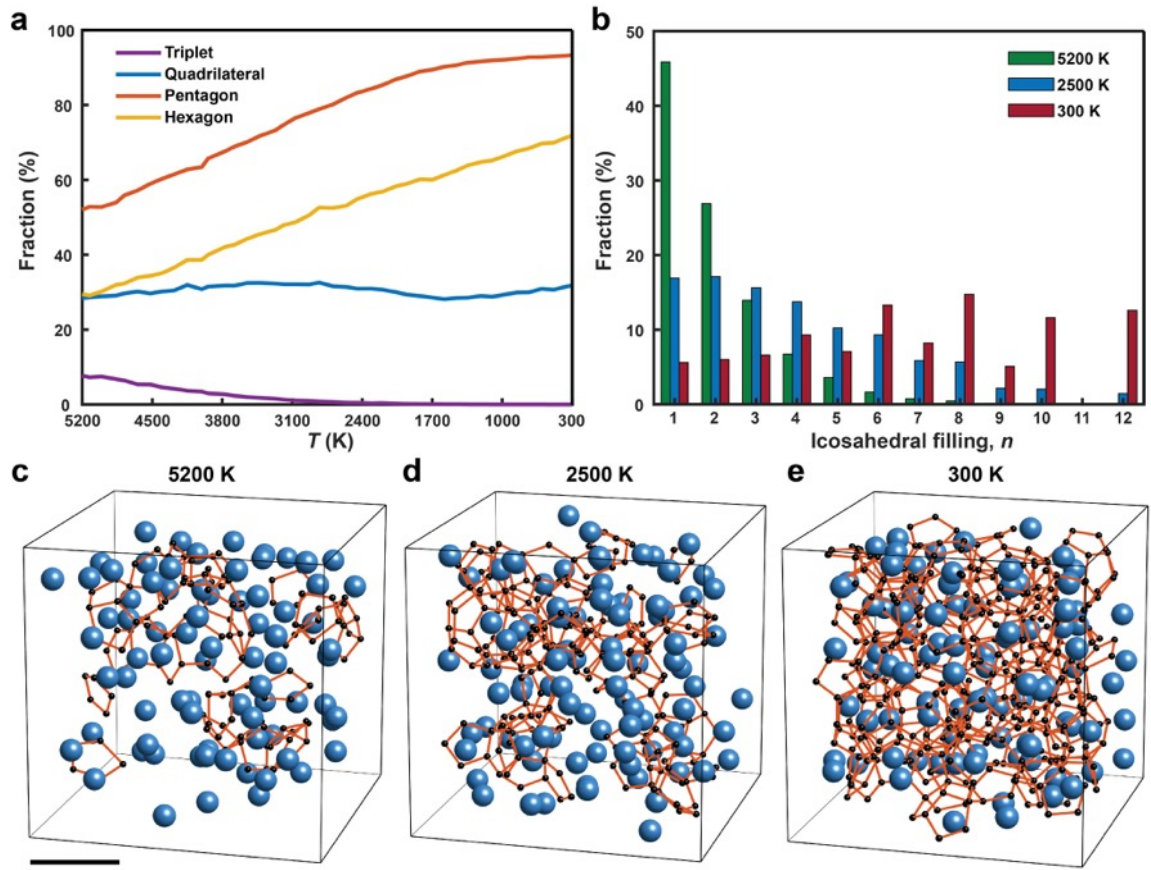

**Fig. 6 | Pentagonal bipyramids networks in the MD simulated Ta liquid and metallic glass. a**, Population of the four most abundant atomic motifs (triplets, quadrilateral, pentagonal and hexagonal bipyramids) as a function of the temperature. **b**, Population of pentagonal bipyramids filling partial icosahedral sites ($n$) at the liquid temperature (5200 K), during quench (2500 K) and at room temperature (300 K). **c-e**, Snapshot of a fraction of the atomic models and PBNs at the above three temperatures. Ta atoms are shown as blue balls and the PBNs as orange lines and black dots. The PBNs quickly grow in size and a fraction of full icosahedra are formed with the decrease of the temperature. Scale bar, 4 Å.

## METHODS

**Physical vapor deposition of amorphous Ta thin films.** Ta thin films were prepared in the ultrahigh vacuum system associated with the SPLEEM (spin polarized low energy electron microscope) instrument at the National Center for Electron Microscopy of the Molecular Foundry at Lawrence Berkeley National

Laboratory. Prior to sample growth, $Si_3N_4$-window grids were cleaned by heating to about 700 K in ultrahigh vacuum. For film depositions, the grids were held in a liquid nitrogen cryostat, where tests using Pt1000 sensors indicated that the temperature of the grids remained in the range of $130 - 150$ K during depositions. High-purity physical vapor beams were produced by heating small charges of Ta in a water-cooled electron beam evaporator and depositing on the $Si_3N_4$-windows to grow amorphous Ta films. The growth chamber has a base pressure in the $10^{-11}$ torr range, and during sample deposition pressure did not exceed the low $10^{-9}$ torr range. Growth rates were calibrated by growing test-films on single crystal Ru(0001), W(110) and Cu(100) substrates while observing low energy electron reflectivity oscillations associated monolayer-by-monolayer epitaxial growth, as well as by measuring peak height ratios in Auger electron spectra of the test samples. All samples were deposited using growth rates in the range of $0.2 - 1.0$ atomic monolayer per minute (or about $2 - 10$ nm/hour). After finishing deposition, all samples were coated with ~2 nm amorphous carbon, deposited from another e-beam evaporator to protect the amorphous structure.

**Synthesis of amorphous Pd nanoparticles.** The amorphous Pd nanoparticles were synthesized by following a previously reported heating-up method with a minor modification[51,52]. Typically, 407 mg palladium (II) acetylacetonate, 4 mL trioctylphosphine and 40 mL oleylamine were placed into a round-neck flask. The mixture was degassed at room temperature for 1 hour under vacuum. The reaction solution was then slowly heated to 280 °C (c.a. 3 ºC/min) under nitrogen, and kept at 280ºC for 30 min. The reaction was quenched by removing the heating mantle and blowing cool air. The product was purified through centrifugation after precipitation with ethanol and the resulting Pd nanoparticles were re-dispersed in toluene. The procedure of ligand exchanges with $NH_2^-$ was as follows[53]: 20 mg of $NaNH_2$ was dissolved in 10 mL dimethyl sulfoxide (DMSO), followed by an addition of 10 mL of the Pd nanoparticle toluene solution (0.4 mg/mL). The mixture was stirred for two days to complete the ligand exchange. The product was collected by centrifugation and washed by acetone one more time. The purified product was obtained by centrifugation and dispersed in nanopore water. Acetone, toluene, DMSO, oleylamine (70%), palladium (II) acetylacetonate (98%), sodium amide ($NaNH_2$, 95%), and trioctylphosphine (97%) were obtained from Sigma Aldrich.

**ADF-STEM data acquisition.** Tomographic tilt series of Ta thin films and Pd nanoparticles were acquired using the TEAM I microscope at the National Center for Electron Microscopy. The microscope was operated in annular dark field mode with electron energy of 300 keV (Supplementary Table 1). A low-exposure acquisition scheme was adopted for our data acquisition[32]. When measuring an image at a tilt angle, a nearby nanoparticle or a feature in the Ta film was used to align and focus the image, thus reducing the unnecessary electron dose to the sample under study. At each tilt angle, three sequential images were taken with a dwell time of 3 μs to minimize the dose rate and drift distortion in each image. To further mitigate the beam damage to the samples, the total electron dose of each tilt series was optimized to be about $8.2$-$9.6 \times 10^5$ $e^-/Å^2$ (Supplementary Table 1). With carefully designed sample preparation and data acquisition protocols, our samples were more stable under the electron beam than some of the previously studied glass samples[54,55]. Images taken at 0º tilt angle before, during and after the tilt series indicate that structural change of the samples throughout the experiment was minimal (Supplementary Fig. 4).

**Electron diffraction experiment and analysis of the amorphous Ta film and Pd nanoparticles.** The electron diffraction patterns of the Ta film and Pd nanoparticles were acquired using a Thermo Fischer Themis transmission electron microscope equipped with a Ceta 2 camera (insets in Supplementary Fig. 5a and c). The accelerating voltage was 300 kV and a 10 μm SA aperture was used to reduce the area, from which the diffraction was collected with the central beam blocked by a beam stop. To calculate the structure factor from each diffraction pattern, a mask was generated to remove the beam stop by properly thresholding the intensity. By fitting the first diffraction ring, the centre of the diffraction pattern was identified and the radially averaged intensity was obtained. A gold sample was used as a reference to calibrate the reciprocal space unit, yielding the intensity distribution as a function of the spatial frequency, I($q$). The structure factor, S($q$), was computed from I($q$) by using the SUePDF software[56], where the atomic form factor was set by properly selecting the chemical species and electron energy. The background was optimized by specifying the pre-peak and the tail location of I($q$), the number of middle reference points and the maximum fitting order. Proper parameters were selected during this step to ensure that the resulting S($q$) oscillates and converges to unity at large $q$ (Supplementary Fig. 5a and c). The PDF was computed by taking the Fourier transform of S($q$). For a diffraction pattern with a high signal-to-noise ratio, its reduced PDF has a linear dependence near the origin, from the slope of which the atomic density can be extracted[41,57]. But our samples are very thin ($\leq$ 10 nm) and their electron diffraction patterns do not have a sufficiently high signal-to-noise ratio, resulting in some oscillations at low spatial frequency. To correct for the oscillations, SUePDF was used to normalize the reduced PDF by fitting a straight line from the origin to the left valley of the first peak, from which the final PDF was obtained (Supplementary Fig. 5b and d).

**Image pre-processing.** The following four steps were used to perform image pre-processing.

i) Drift correction[34]. To correct sample drift, three ADF-STEM images at each tilt angle were registered using the following procedure. First, a region of 400 × 400 pixels from the second and third images was cropped and scanned over the first one with a step size of 0.05 pixel. Next, the cross-correlation coefficient between images was calculated, where the relative drift vectors were identified by the maximum cross-correlation. The ADF-STEM images have a typical drift of less than 1 pixel. We then applied drift distortion correction to each image along the slow scan direction. By assuming a linear drift during the data acquisition, the drift for each pixel in the image can be determined and corrected by interpolating the raw image with drift corrected pixel positions. Finally, the three drift corrected images at each tilt angle were averaged for denoising.

ii) Image denoising. The experimental images have both Poisson and Gaussian noise. A general algorithm named Block-matching and 3D filtering (BM3D)[58] has been proven effective in previous AET experiments[34,36,37], and was applied to denoise the drift corrected images. To optimize the BM3D parameters, we first estimated the Gaussian and Poisson noise in the experimental images, and then applied BM3D to denoise a simulated ADF-STEM image with the same noise level by varying denoising parameters. The best parameters were identified by maximizing the cross-correlation between the denoised image and the simulated noise-free image. These optimized parameters were used to denoise all the experimental images.

iii) Background subtraction. For each denoised image, a mask slightly larger than the sample was generated by thresholding. From the background outside the mask, the background level within the mask was estimated using Laplacian interpolation. The estimated background was subtracted from the denoised image.

iv) Image alignment. The images in each tilt series were aligned using the following procedure. The tilt series of the two Pd nanoparticles was aligned by the centre of mass and common line method, as described in previous AET experiments[31,32]. For the Ta thin film, we first performed a pre-alignment by using cross-correlation between the images of neighbouring tilt angles. Next, based on reference markers in the sample (in this case we chose an isolated region as the reference marker), we used the common line method and the centre of mass[31,32] to align the thin film along the tilt axis and perpendicular to the tilt axis, respectively. We repeated this alignment process until no further improvement could be made.

**3D reconstruction with RESIRE.** After image pre-processing, each experimental tilt series was computed by the Real Space Iterative Reconstruction (RESIRE) algorithm. The algorithm minimizes the difference between the experimental and computed images using the gradient descent method. The error function and the gradient are defined as,

$$\varepsilon_\theta(O) = \frac{1}{2}\sum_{x,y}|\Pi_\theta(O)\{x,y\} - b_\theta\{x,y\}|^2 \qquad (1)$$

$$\nabla\varepsilon_\theta(O)\{u,v,w\} = \Pi_\theta(O)\{x,y\} - b_\theta\{x,y\} \quad where \begin{bmatrix} u \\ v \\ w \end{bmatrix} = R_\theta \begin{bmatrix} x \\ y \\ z \end{bmatrix} \; for \; some \; z \quad (2)$$

where $\varepsilon_\theta$ is the error function of a 3D object $O$ at tilt angle $\theta$, $\Pi_\theta(O)$ calculates the projection of the object $O$ at angle $\theta$, $b_\theta$ is the experimental image at angle $\theta$, $\{x,y,z\}$ are the coordinates, $\nabla$ is the gradient, $R_\theta$ is the rotation matrix transforming coordinates $\{x,y,z\}$ to $\{u,v,w\}$. More mathematical description of RESIRE and its superior performance to other algorithms, such as the generalized Fourier iterative reconstruction and the simultaneous iterations reconstruction technique, will be presented in a follow-up paper.

The two Pd nanoparticles were directly reconstructed by RESIRE. For the Ta thin film, we first performed a large volume reconstruction with RESIRE. After estimating the thickness variation of the Ta thin film, we applied scanning AET to do multiple local volume reconstructions and then patched them together to obtain a full reconstruction[38]. Previous study has shown that scanning AET can improve the reconstruction of 2D layered and thin film samples over AET[38].

From the 3D reconstructions of the Pd nanoparticles and Ta film, we performed angular refinement and image alignment until the results converged. Next, we traced and refined the 3D atomic coordinates from each reconstruction (see the section below), from which reference images were calculated at the corresponding tilt angles. The experimental tilt angles and images were further refined and re-aligned using these reference images. After these procedures, the final reconstructions were performed for all three samples.

**Determination and refinement of 3D atomic coordinates.** From the reconstructions of the Pd nanoparticles and the Ta thin film, we determined their 3D atomic coordinates using the following steps.

i) We interpolated each reconstruction onto a finer grid with three times oversampling using the spline method. All the local maxima in the reconstruction were identified and the positions of potential atoms were extracted from a local volume of 0.8 Å × 0.8 Å × 0.8 Å with a polynomial fitting method[37,59]. For every potential atom, a minimum distance of 2 Å to its neighbouring atoms have to be satisfied. This constraint is based on the fact that all the interatomic distances in our samples are larger than 2 Å. After iterating through all local maxima, a list of potential atoms was obtained.

ii) To separate non-atoms from the potential atoms, we employed the K-mean clustering method[36,37,60] based on the integrated intensity of a local volume (0.8 Å × 0.8 Å × 0.8 Å) around each potential atom position. After excluding non-atoms, the potential atomic models of the two Pd nanoparticles and Ta film were obtained.

iii) By carefully comparing the individual atomic positions in the potential atomic models with the 3D reconstructions, we manually corrected a very small fraction (typically < 1%) of unidentified or misidentified atoms. Note that manual correction of a very small fraction of atoms is routinely used in macromolecular crystallography for atom tracing and refinement[61].

iv) We repeated step ii) and iii) until no further improvement could be made, resulting in the 3D atomic models of the three amorphous materials.

v) The 3D atomic coordinates in each model were refined by minimizing the error between the experimental and computed images using the gradient descent as described elsewhere[34,36,37]. The refinement results are shown in Supplementary Table 1.

**3D precision estimation.** To estimate the 3D precision of our method, we performed multislice simulations[62,63] to calculate ADF-STEM images from the Ta atomic model using the same experimental parameters specified in Supplementary Table 1. A tilt series of 46 multislice images was computed at the experimental tilt angles. To account for the electron probe size and other effect, the image was convolved with a Gaussian function. A representative multislice ADF-STEM image at 0° tilt angle is shown in Supplementary Fig. 6b, which is in good agreement with the corresponding experimental image (Supplementary Fig. 6a). From the 46 multislice ADF-STEM images, we used the same reconstruction, tracing and refinement procedure to obtain a new 3D atomic model. By comparing the new atomic model with the experimental one, we find 98.1% of the atoms are identical with a root-mean-square deviation of 18 pm (Supplementary Fig. 6c).

**Calculation of the pair distribution function from the experimental 3D atomic model.** The following procedure was used to calculate the PDF from the experimental atomic model of each sample[57]. i) The histogram of atom pair distances in spherical shells was computed with a shell thickness of 0.1 Å. ii) The counts in each spherical shell were divided by the volume of the spherical shell, yielding the density of atom pairs as a function of the pair distance. iii) The PDF was scaled to approach one at large pair distances. Using this procedure, we calculated the PDFs of all the atoms in the Ta thin film and two Pd nanoparticles. From the PDF of each material, we determined the first valley position, corresponding to the first nearest neighbour shell distance. This distance was used to compute the local bond orientational order (BOO) parameters (see the section below), from which crystal nuclei were identified. After excluding the crystal nuclei, the PDFs of the disordered atoms in the amorphous materials were re-calculated, shown in Fig. 1c.

**The local bond orientational order parameters.** We calculated the averaged local BOO parameters ($Q_4$ and $Q_6$) to quantify the disorder of the amorphous materials[42,64]. The $Q_4$ and $Q_6$ values were computed based on the procedure published elsewhere[42], where the first nearest neighbour shell distance (Fig. 1c) was used as a constraint. $Q_4$ and $Q_6$ were used to calculate the normalized BOO parameter, defined as $\sqrt{Q_4{}^2 + Q_6{}^2}/\sqrt{Q_{4\,\text{fcc}}^2 + Q_{6\,\text{fcc}}^2}$, where $Q_{4\,\text{fcc}}$ and $Q_{6\,\text{fcc}}$ are the reference values of the fcc lattice. We separated crystal nuclei from amorphous structure by setting the normalized BOO parameter larger than or equal to 0.5[40] (red dashed lines in Supplementary Fig. 8a-c).

**The Voronoi tessellation, coordination number and local mass density distribution.** The Voronoi tessellation of each 3D atomic model was calculated by following procedure published elsewhere[4]. To characterize the nearest neighbour atoms around each centre atom, a regulation was applied to each Voronoi polyhedron, where neighbouring atoms with the facet area less than 1% of the total Voronoi surface area were removed during the analysis[23]. All the Voronoi polyhedra were then indexed by $\langle n_3, n_4, n_5, n_6 \rangle$ with $n_i$ denoting the number of $i$-edged faces. The coordination number was calculated by $\sum_i n_i$.

The mass density for each atom was calculated by dividing the atomic mass by its atomic volume, which is defined as the volume of its Voronoi polyhedron without regulation. The densities at all atomic positions were interpolated onto a 3D grid and then convolved with a Gaussian kernel. The width of the Gaussian kernel was set as the first nearest neighbour shell distance defined by the PDF. Using this procedure, we obtained the local mass density distribution of the three amorphous materials (Fig. 3 and Supplementary Fig. 10).

**Polytetrahedral packing analysis.** To identify all the tetrahedra in each amorphous material, we used the distortion parameter ($\delta$) defined in the main text, where the maximum edge length of each tetrahedron cannot be larger than the first nearest neighbour shell distance (Fig. 1c). In the polytetrahedral packing analysis, we set $\delta$ to be the maximum allowed distortion parameter to identify each tetrahedron. The population of the tetrahedra strongly depends on $\delta$. With $\delta > 0.2$, more than 96.8% of the atoms in the three amorphous materials form tetrahedra (green curves in Supplementary Fig. 9). From these tetrahedra, we searched for polytetrahedral motifs and found four main motifs (Fig. 2a): i) triplets with three face-sharing tetrahedra (but the 1[st] and 3[rd] tetrahedron do not share a face); ii) quadrilateral bipyramids with four face-sharing tetrahedra; iii) pentagonal bipyramids with five face-sharing tetrahedra; and iv) hexagonal bipyramids with six face-sharing tetrahedra. All these four polytetrahedral motifs share two capping atoms (brown atoms in Fig. 2a). Although we observe other motifs in the amorphous materials, their population is much smaller than the four main motifs. We represented the four main motifs by three-, four-, five- and six-fold skeletons, which connect the centroids of the tetrahedra (Fig. 2a). The fraction of four main motifs, defined as the number of tetrahedra in each motif divided by the total number of tetrahedra in each amorphous material, strongly depends on $\delta$ (Supplementary Fig. 9). The sum of the fraction of four main motifs can be larger than 1 because some tetrahedra are overlapped among different motifs. By choosing $\delta \leq 0.255$[14,44], we find the polytetrahedral packing of the four motifs is strongly correlated with the local mass density heterogeneity (Fig. 3 and Supplementary Fig. 10).

**Pentagonal bipyramid networks**. We searched for the PBNs in each amorphous material using the following procedure. i) From the polytetrahedral packing of the sample, we only kept pentagonal bipyramids, which are represented by five-fold skeletons. ii) We started a PBN by choosing a five-fold skeleton and identifying all its edge-sharing skeletons. iii) We repeated step ii) until all the edge-sharing skeletons in the PBN were found. iv) We started a new PBN and repeated the procedure. v) After identifying all the PBNs in the sample, we only kept those PBNs with five or more pentagonal bipyramids. The five largest PBNs in the Ta thin film, two Pd nanoparticles and MD simulated Ta liquid (5200K) are shown in Supplementary Figs. 16-18 and 20. The PBN size is defined as the number of pentagonal bipyramids in the network. The PBN length was measured along the longest direction of the network (Supplementary Fig. 15). We used the same procedure to find the quadrilateral and hexagonal bipyramid networks in each amorphous material.

**Molecular dynamics simulations.** To understand our experimental observations, we performed MD simulations of a Ta bulk system using the LAMMPS package[45]. The system consisted of 31250 atoms using the embedded-atom-method interatomic potential[18] with periodic boundary conditions. The system was melted and equilibrated at 5200 K before quenching the system at a cooling rate of $10^{13}$ K/s using the isothermal-isobaric ensemble. The polytetrahedral packing analysis was performed on atomic configurations throughout the quench process and we observed similarities between the experimental amorphous materials and the MD simulations of liquids. This realistic interatomic potential was chosen for three reasons. First, the interatomic potential was developed with a focus on the metallic glass and liquid phases. Second, the simulated metallic glass structure factors correlate well with the experimental data[18]. Third, the simulated liquid phase PDF has the lowest least-square fitting error to the experimental Ta PDF (Fig. 1c) compared to other potentials[65,66], although all these interatomic potentials show similar PBNs and trends through the glass transition.

**Quantifying the crystalline-amorphous interface.** We quantified the characteristic width of the crystalline-amorphous interface using the normalized BOO parameter. In each sample, the 3D shape of every nucleus was determined by setting the normalized BOO parameter larger than or equal to 0.5. The perpendicular distance of all atoms to the surface of the 3D shape was calculated, where the atoms outside and inside the surface have a positive and negative distance, respectively. By averaging all the nuclei in each sample, we obtained the distribution of the normalized BOO parameter as a function of the perpendicular distance to the surface of the crystal nuclei (blue circles in Supplementary Fig. 22). To extract the characteristic width ($d_c$) of the crystalline-amorphous interface, the experimental data points were fitted with an exponential decay function $y = ae^{-x/d_c} + b$ (solids curves in Supplementary Fig. 22), where $a$ and $b$ are two constants. The characteristic width of the crystalline-amorphous interface was determined to be 3.0, 4.2 and 4.3 Å for the Ta film, $Pd_1$ and $Pd_2$ nanoparticles, respectively, which agree with previous MD simulated results[46].

## Supplementary Table and Figures

**Supplementary Table 1 | AET data collection, processing, reconstruction, refinement and statistics**

|  | Ta film | Pd$_1$ nanoparticle | Pd$_2$ nanoparticle |
|---|---|---|---|
| **Data Collection and Processing** |  |  |  |
| Voltage (kV) | 300 | 300 | 300 |
| Convergence semi-angle (mrad) | 17.1 | 17.1 | 17.1 |
| Probe size (Å) | 0.7 | 0.7 | 0.7 |
| Detector inner angle (mrad) | 30 | 30 | 30 |
| Detector outer angle (mrad) | 195 | 195 | 195 |
| Depth of focus (nm) | 14 | 14 | 14 |
| Pixel size (Å) | 0.322 | 0.454 | 0.454 |
| Number of projections | 46 | 52 | 54 |
| Tilt range ($^{o}$) | -63.4 | -72.0 | -72.4 |
|  | +55.5 | +66.4 | +69.4 |
| Electron dose ($10^5$ e/Å$^2$) | 8.2 | 9.3 | 9.6 |
| **Reconstruction** |  |  |  |
| Algorithm | RESIRE | RESIRE | RESIRE |
| Oversampling ratio | 3 | 4 | 3.5 |
| Number of iterations | 500 | 200 | 200 |
| **Refinement** |  |  |  |
| R$_1$ (%)[a] | 15.5 | 8.5 | 6.9 |
| R (%)[b] | 14.2 | 5.2 | 4.5 |
| B' factors (Å$^2$) | 22.1 | 45.2 | 58.3 |
| Number of atoms | 8284 | 52308 | 76238 |

[a]The R$_1$-factor is defined by Eq. (5) in ref. 36.

[b]The R-factor is defined by $R = \frac{1}{n}\sum_{\theta}\frac{\sum_{x,y}|\Pi_{\theta}(O)\{x,y\}-b_{\theta}\{x,y\}|}{\sum_{x,y}|b_{\theta}\{x,y\}|}$, where $n$ is the number of projection images.

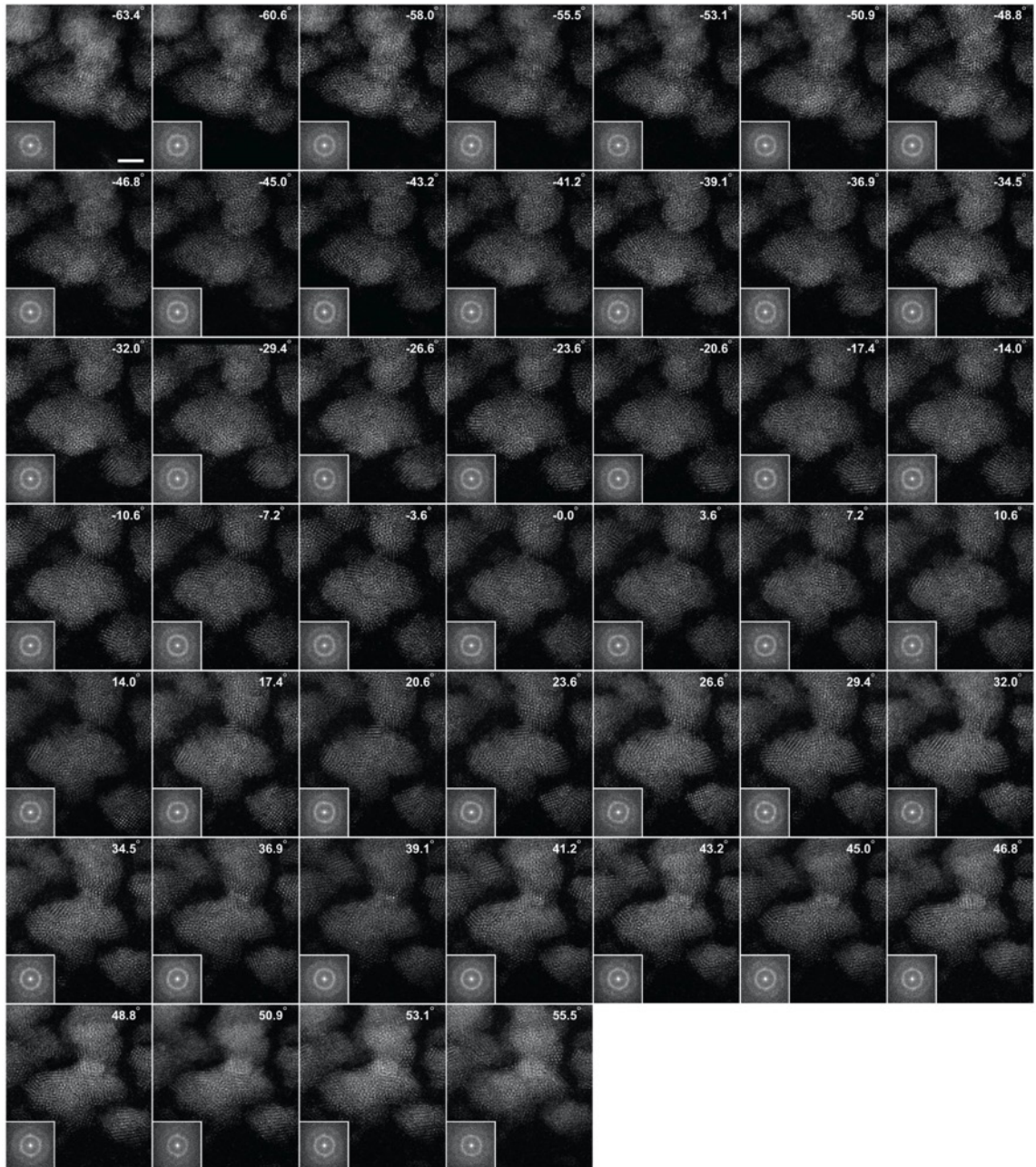

**Supplementary Fig. 1 | Tomographic tilt series of an amorphous Ta thin film.** 46 ADF-STEM images of the Ta film with a tilt range from −63.4° to +55.5°. The insets show the 2D power spectra of the experimental images (in a logarithmic scale), where the amorphous halo is visible. Scale bar, 2 nm.

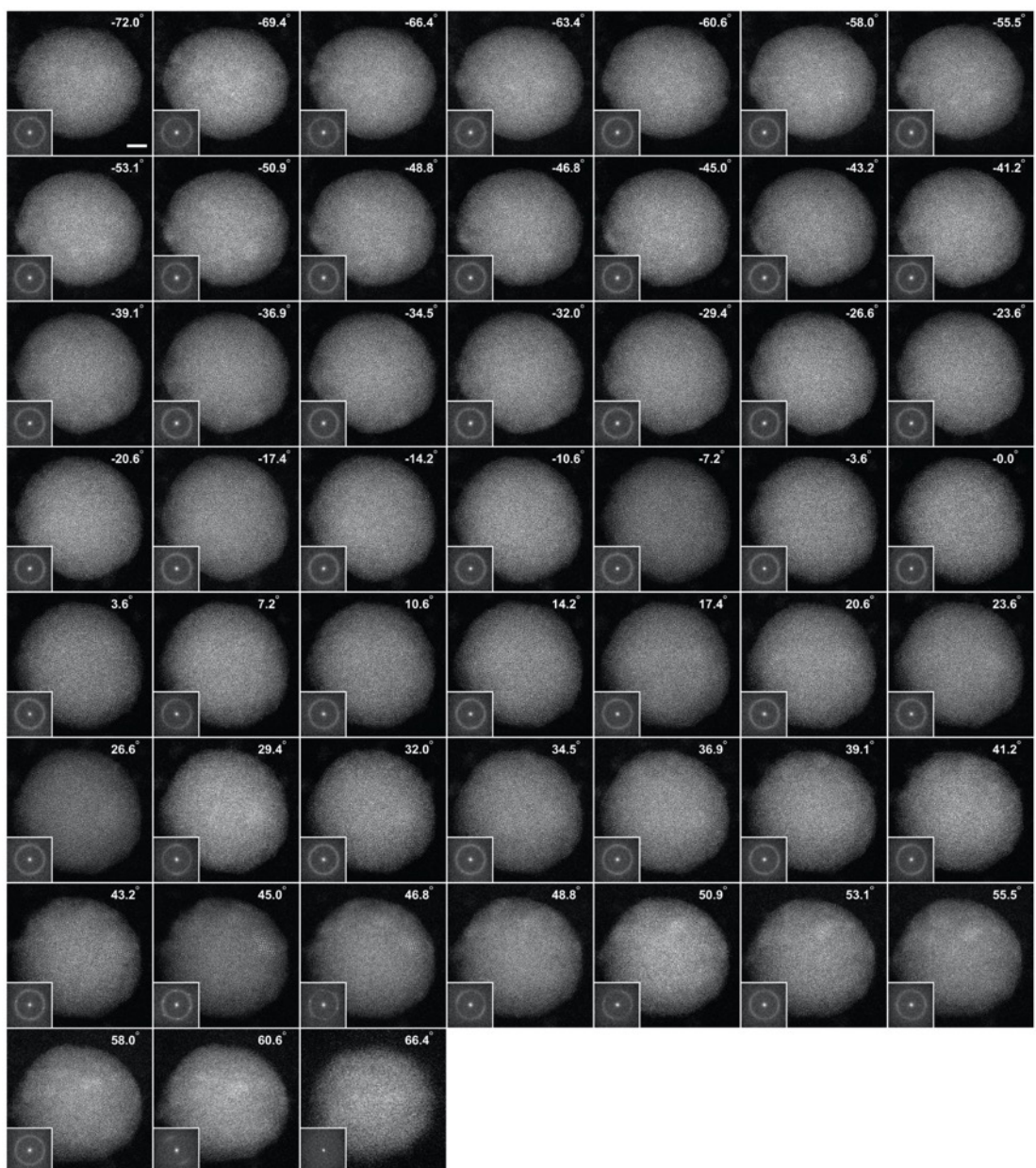

**Supplementary Fig. 2 | Tomographic tilt series of an amorphous Pd nanoparticle (Pd₁).** 52 ADF-STEM images of the Pd₁ nanoparticle with a tilt range from −72.0° to +66.4°. The insets show the 2D power spectra of the experimental images, where the amorphous halo is visible. Scale bar, 2 nm.

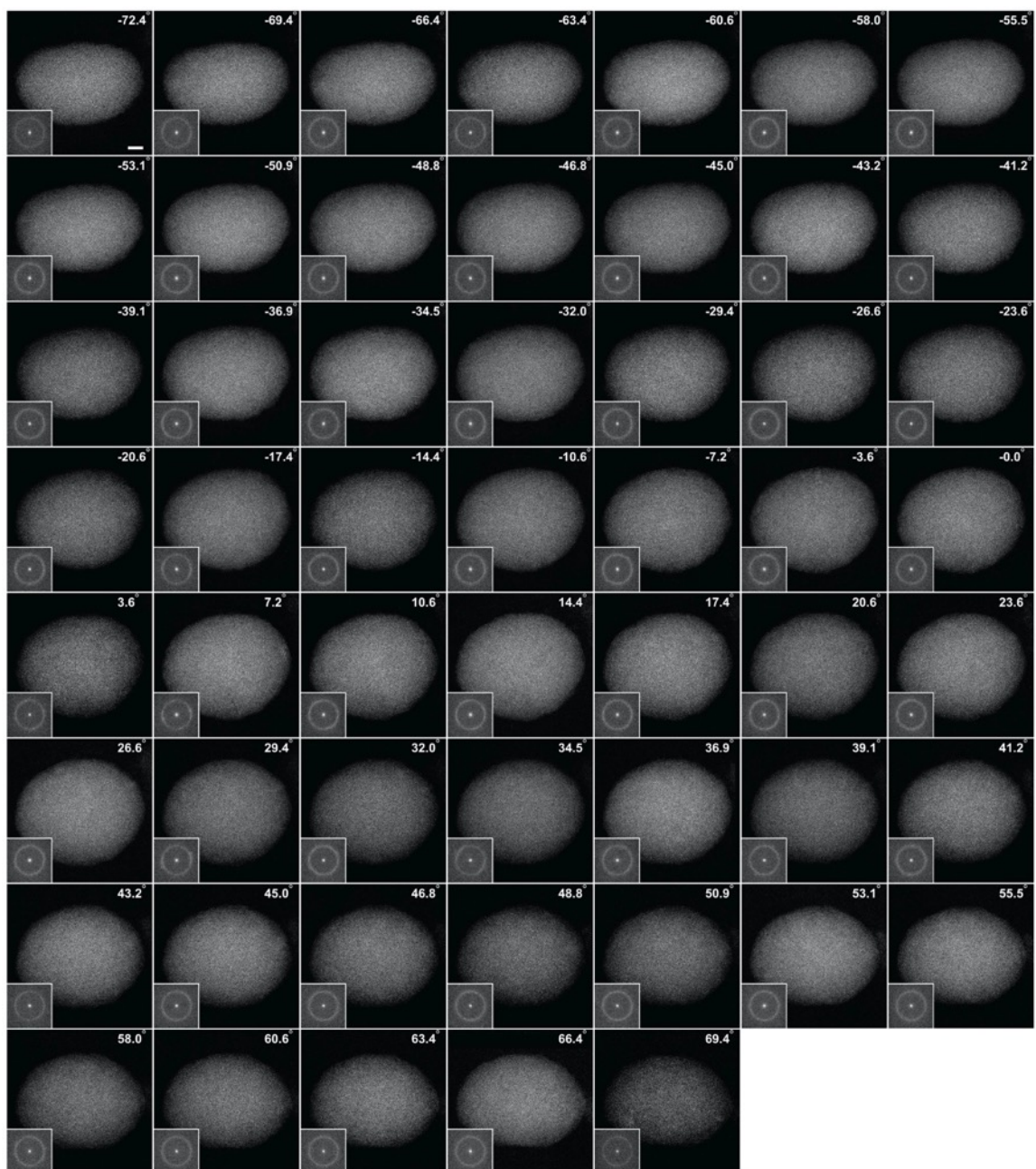

**Supplementary Fig. 3 | Tomographic tilt series of an amorphous Pd nanoparticle (Pd₂).** 54 ADF-STEM images of the Pd₂ nanoparticle with a tilt range from −72.0° to +69.4°. The insets show the 2D power spectra of the experimental images, where the amorphous halo is visible. Scale bar, 2 nm.

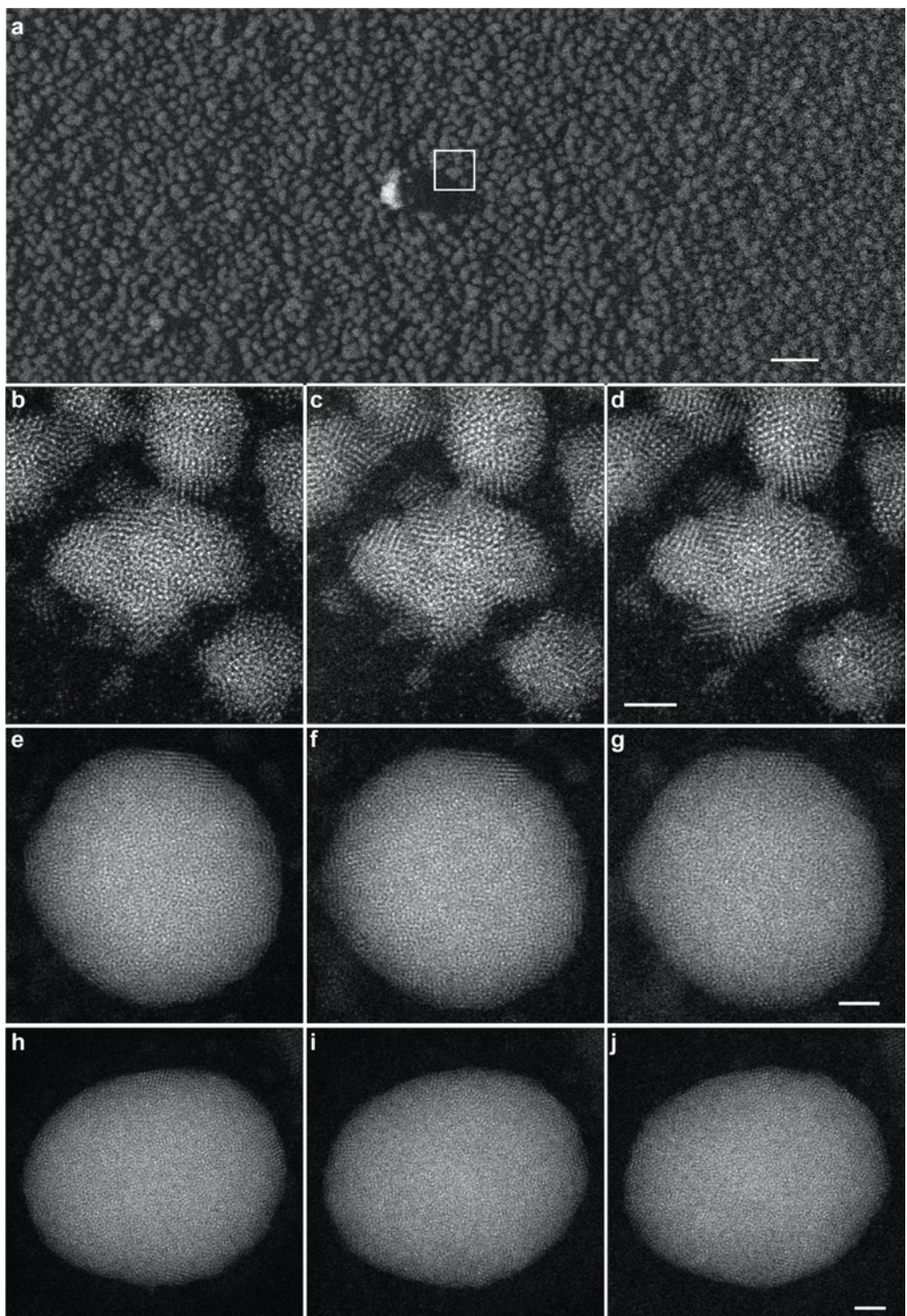

**Supplementary Fig. 4 | Overview of the amorphous Ta thin film and consistency check of the sample structure. a**, Low magnification overview of the Ta thin film. A tomographic tilt series was acquired from the region marked by the white square. ADF-STEM images taken before, during and after the experiment for the Ta thin film (**b-d**), Pd$_1$ (**e-g**) and Pd$_2$ (**h-j**) nanoparticles, indicating a minimal change of the sample structure during the data acquisition. Scale bar, 20 nm in (**a**) and 2 nm in (**d, g, j**). All the images were taken at 0° tilt angle.

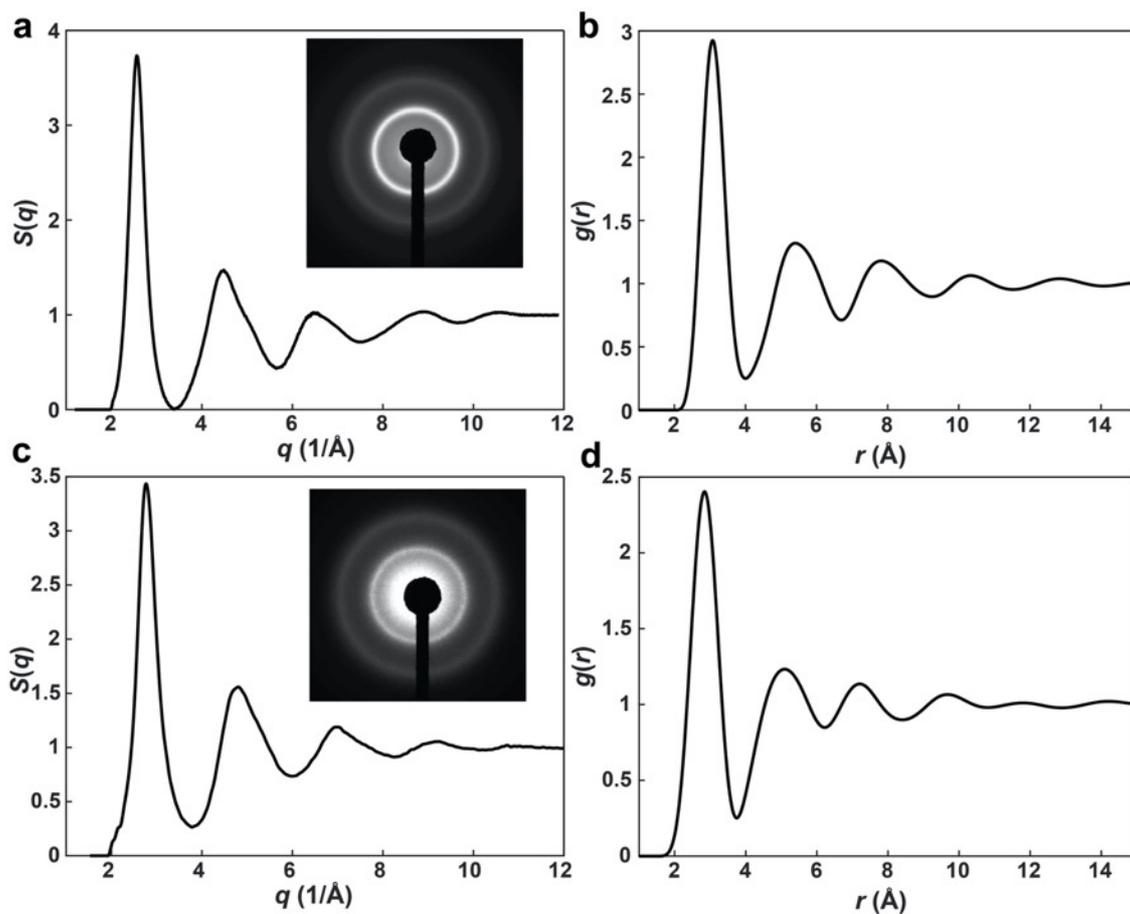

**Supplementary Fig. 5 | Electron diffraction experiment and analysis. a**, Structure factor *S(q)* of the amorphous Ta thin film, obtained from the experimental diffraction pattern (inset). **b**, PDF of the Ta film derived from (**a**). **c**, Structure factor *S(q)* of the amorphous Pd nanoparticles, obtained from the experimental diffraction pattern (inset). **d**, PDF of the Pd nanoparticles derived from (**c**).

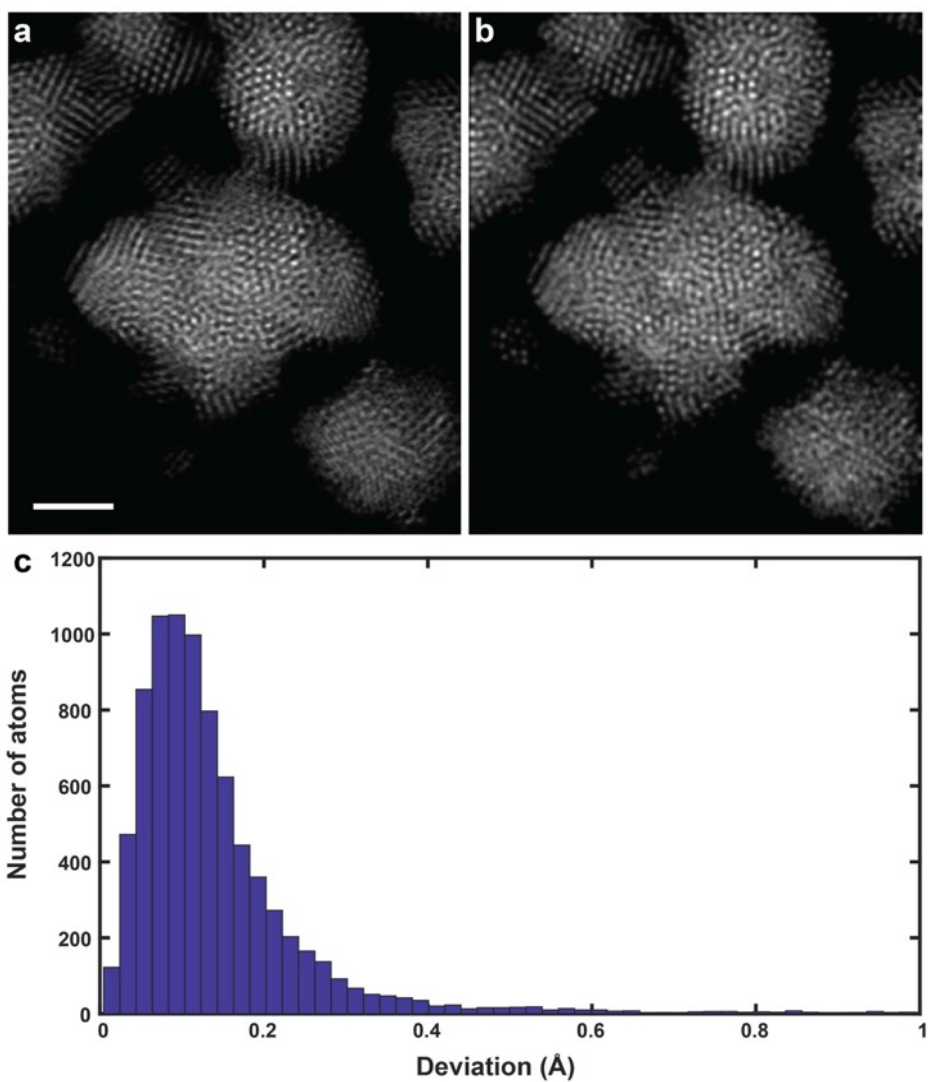

**Supplementary Fig. 6 | 3D precision estimation of the experimental 3D atomic model. a**, Experimental image of the amorphous Ta film at 0° after denoising. **b**, Multislice simulation image calculated from the experimental 3D atomic model of the Ta thin film, which is in good agreement with (**a**). By using 46 multislice simulation images and the same reconstruction, atom tracing and refinement procedure, we obtained a new 3D atomic model. **c**, Histogram of the root-mean-square deviation between the experimental 3D atomic model and the new 3D atomic model, showing 98.1% of the atoms are identical with a 3D precision of 18 pm. Scale bar, 2 nm.

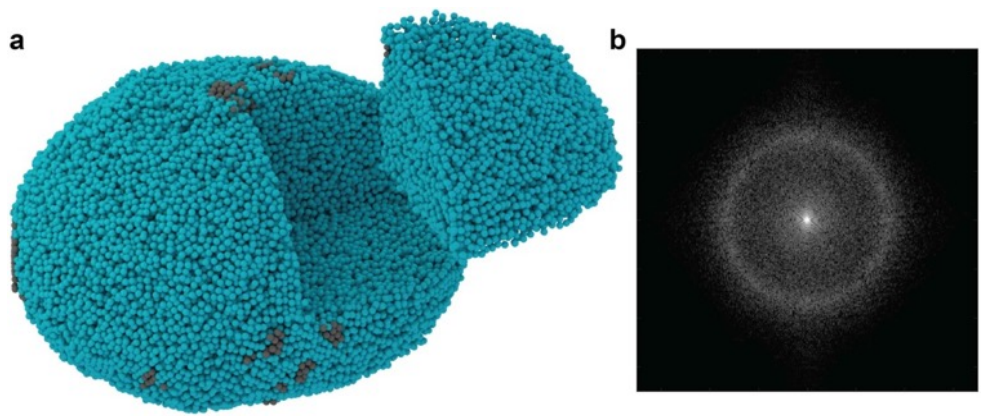

**Supplementary Fig. 7 | Experimental 3D atomic structures of the amorphous Pd₂ nanoparticle**. **a**, Experimental 3D atomic model of the Pd₂ nanoparticle with crystal nuclei in grey. Scale bar, 2 nm. **b**, Average 2D power spectrum of the experimental images of the nanoparticle shows the amorphous halo.

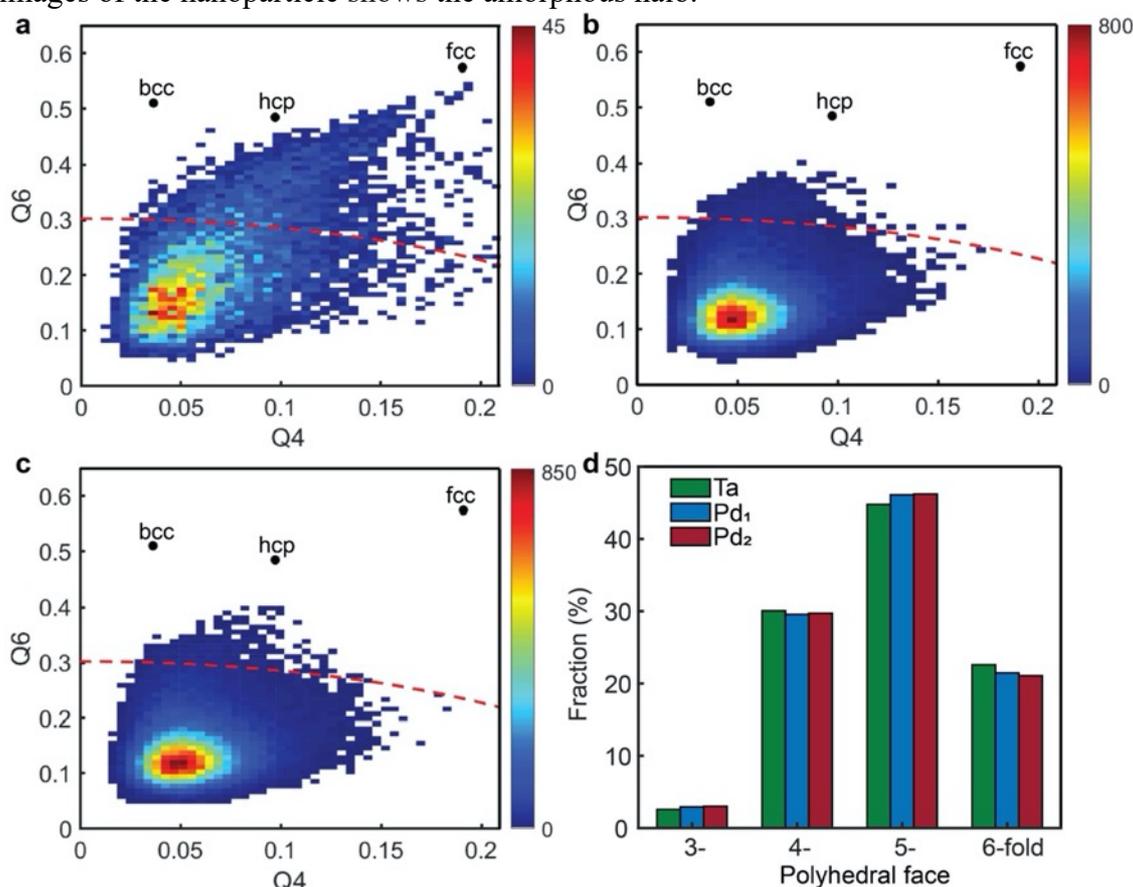

**Supplementary Fig. 8 | Local Bond orientational order parameters for the Ta thin film and two Pd nanoparticles.** Averaged local bond orientational order parameters for the Ta film (**a**), Pd₁ (**b**) and Pd₂ (**c**) nanoparticles. Red dashed lines correspond to the normalized bond orientational order parameter equal to 0.5. **d**, The three-, four-, five- and six-edged face distribution for the Voronoi polyhedra in three samples, where the five-edge faces are the most abundant in all three samples.

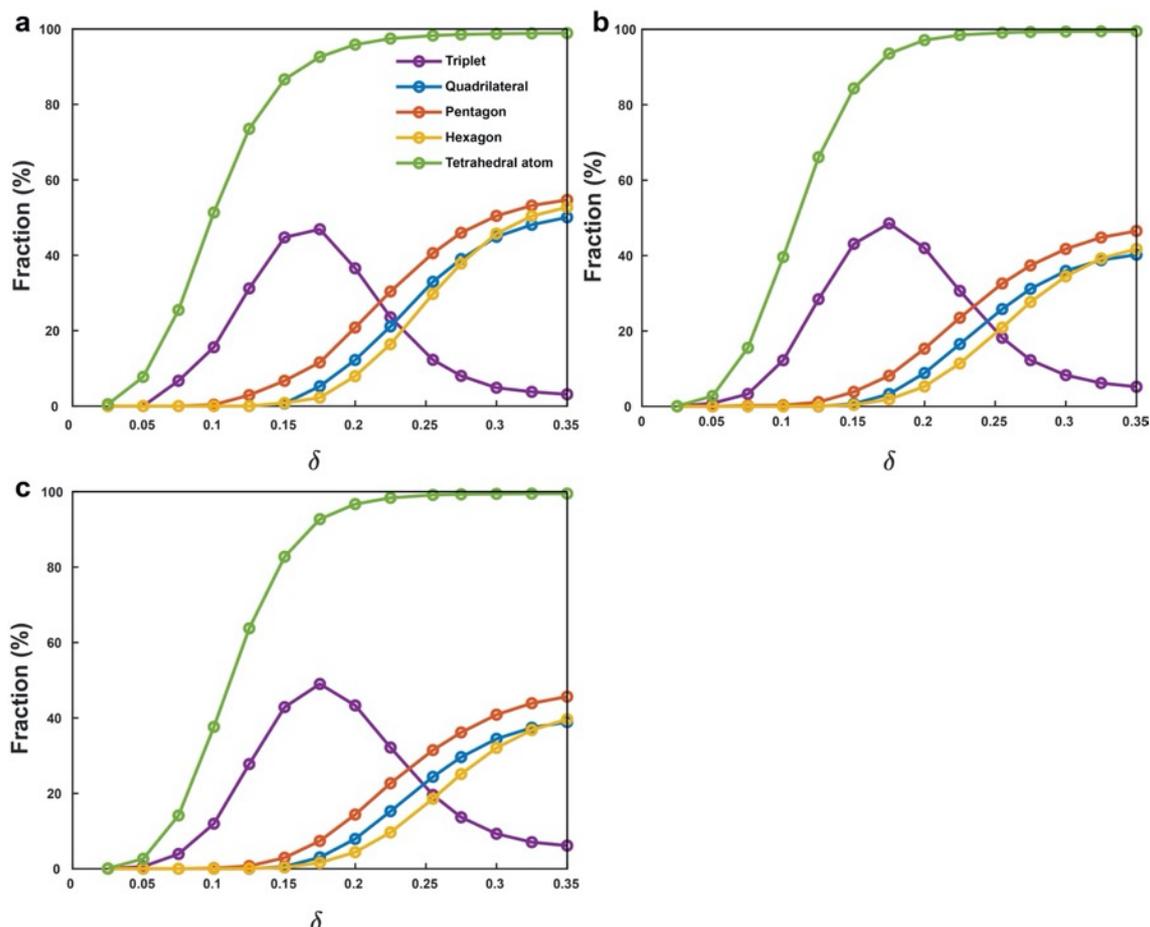

**Supplementary Fig. 9 | Population of tetrahedral atoms and polytetrahedral motifs**. Fraction of the tetrahedral atoms and four polytetrahedral motifs (triplets, quadrilateral, pentagonal and hexagonal bipyramids) in the amorphous Ta film (**a**), Pd$_1$ (**b**) and Pd$_2$ (**c**) nanoparticle, as a function of $\delta$. The green curves show the fractions of the atoms in the three amorphous materials forming tetrahedra. The purple, blue, orange and yellow curves show the fractions of four main motifs, defined as the number of tetrahedra in each motif divided by the total number of tetrahedra in each amorphous material.

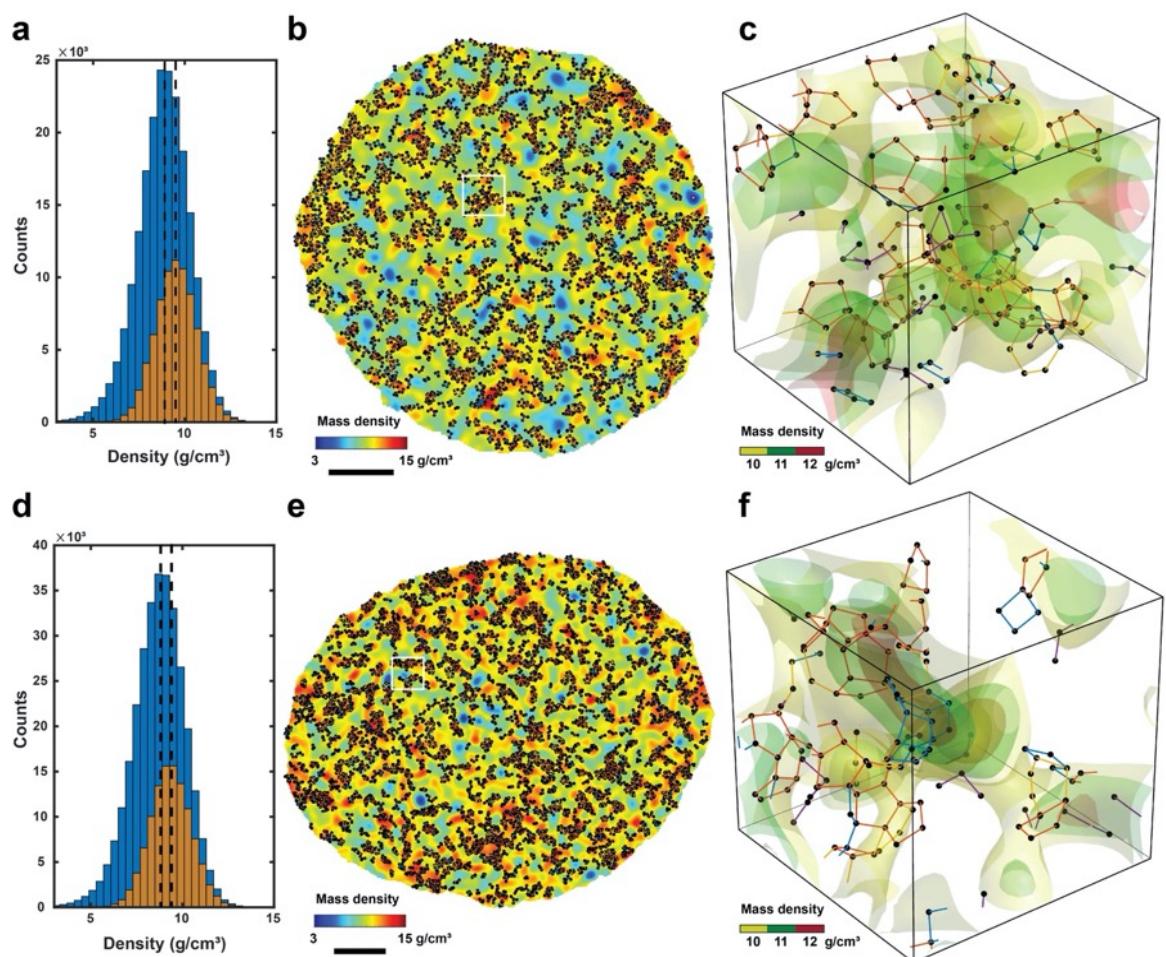

**Supplementary Fig. 10 | Correlation of 3D local mass density heterogeneity and polytetrahedral packing.** Mass density histograms of the amorphous $Pd_1$ (**a**) and $Pd_2$ nanoparticle (**d**) with (yellow) and without polytetrahedral packing (blue), where polytetrahedral packing increases the average mass density by 6.7% and 6.3%, respectively. Slices through the $Pd_1$ (**b**) and $Pd_2$ nanoparticles (**e**) show the local mass density heterogeneity (colour) overlaid with polytetrahedral packing (black). **c**, **f**, 3D surface renderings of local mass density heterogeneity magnified from the square regions in (**b**) and (**e**), respectively, which are overlaid with three- (purple), four- (blue), five- (orange) and six-fold (yellow) skeletons. Scale bar, 2 nm.

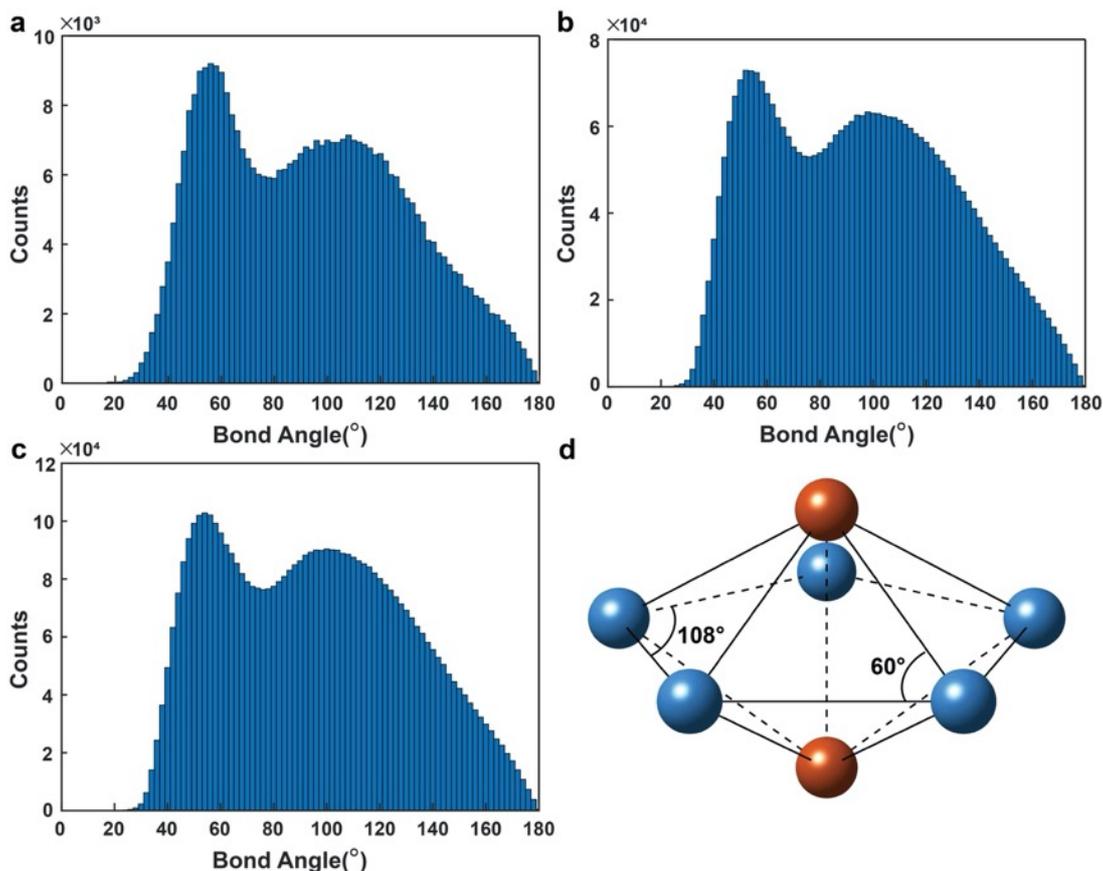

**Supplementary Fig. 11 | Bond angle distributions of the amorphous Ta film (a), Pd₁ (b) and Pd₂ (c) nanoparticles**, which agree with the previous study of liquid metals by combining experimental measurements with reverse Monte Carlo simulations[12]. The two peaks of the bond angle distributions are consistent with the internal angles of a tetrahedron and a pentagon (**d**).

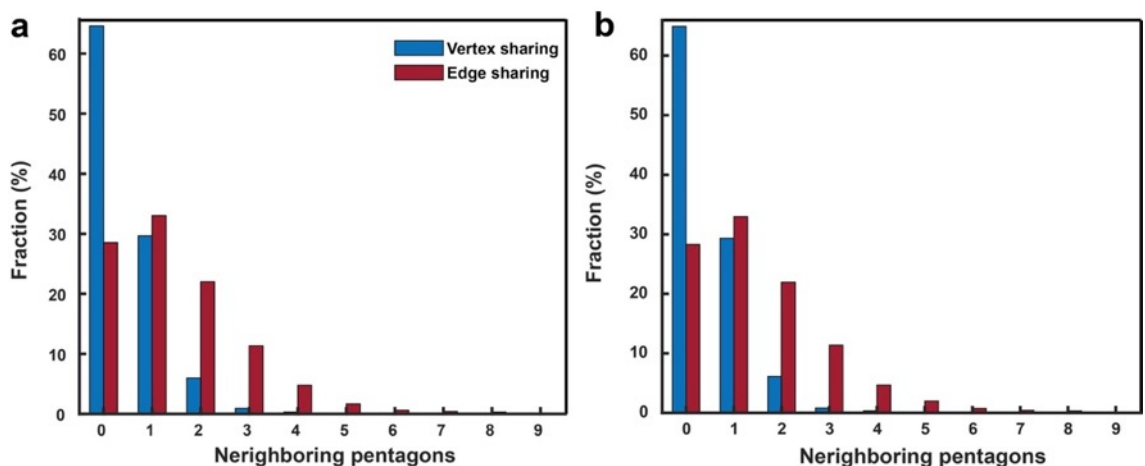

**Supplementary Fig. 12 | Vertex- and edge-sharing the five-fold skeletons.** Population of pentagonal bipyramids as a function of the number of vertex- (**blue**) and edge-sharing (red) neighbours in the amorphous Pd₁ (**a**) and Pd₂ (**b**) nanoparticle. Edge-sharing of the five-fold skeletons have more neighbours than vertex-sharing of five-fold skeletons.

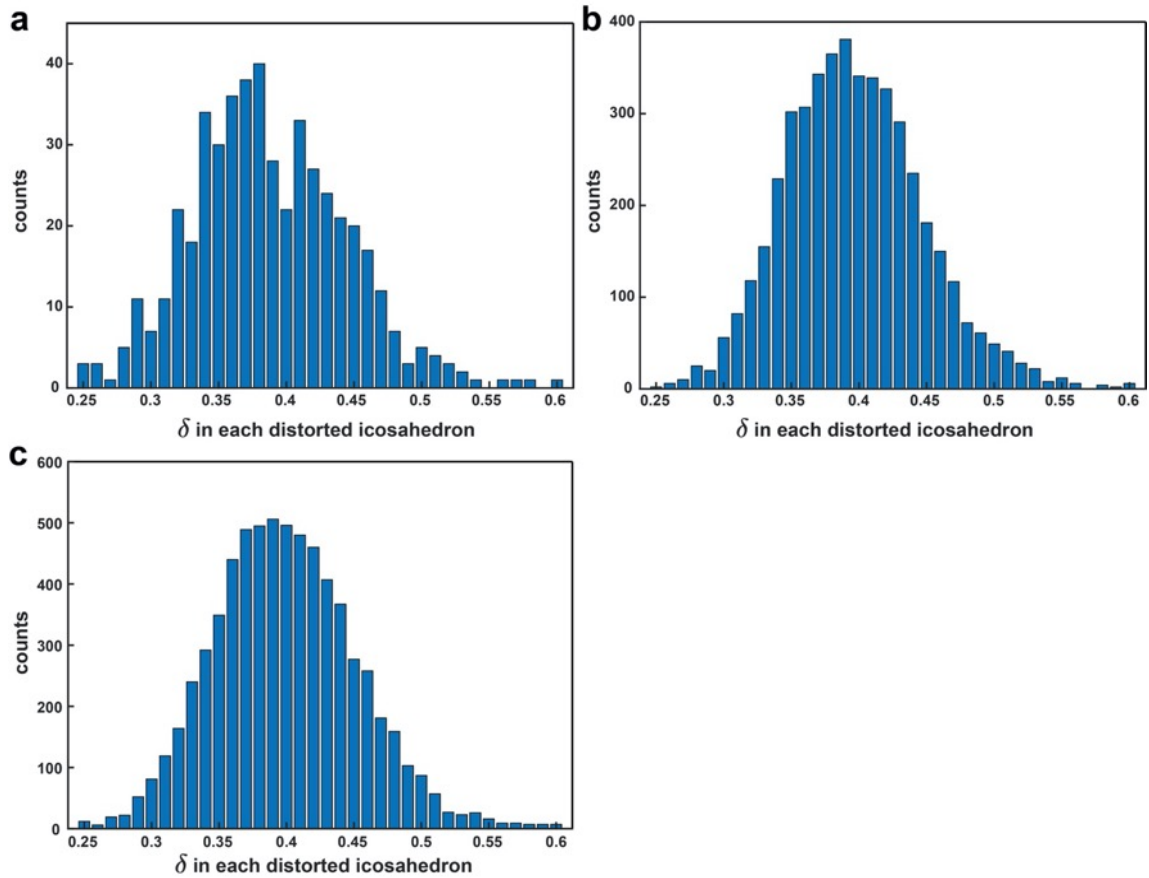

**Supplementary Fig. 13 | Distorted icosahedra in Voronoi tessellation.** Number of icosahedra (counts) with <0,0,12,0>, <0,1,10,2>, <0,2,8,2> and <0,2,8,1> as a function of $\delta$ in the amorphous Ta film (**a**), Pd$_1$ (**b**) and Pd$_2$ (**c**) nanoparticles, where the vast majority of the distorted icosahedra in the Voronoi tessellation have a large distortion with $\delta > 0.255$. When choosing $\delta \leq 0.255$, there are only 17 distorted icosahedra in the three amorphous materials.

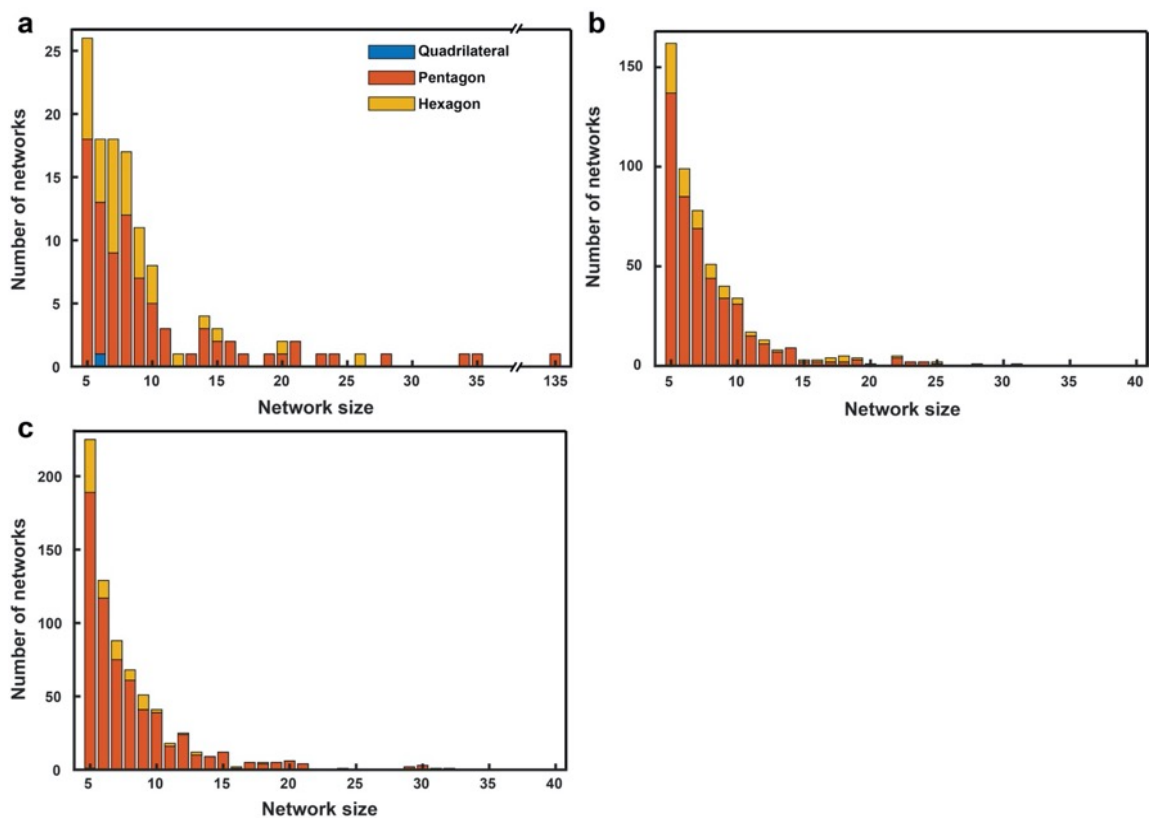

**Supplementary Fig. 14 | Analysis of the quadrilateral, pentagonal, and hexagonal bipyramid network size.** Population of the quadrilateral (blue), pentagonal (orange), and hexagonal bipyramid (yellow) networks as a function of their size in the amorphous Ta thin film (**a**), Pd$_1$ (**b**) and Pd$_2$ (**c**) nanoparticle, where the network size is defined as the number of quadrilateral, pentagonal, and hexagonal bipyramids in the corresponding networks.

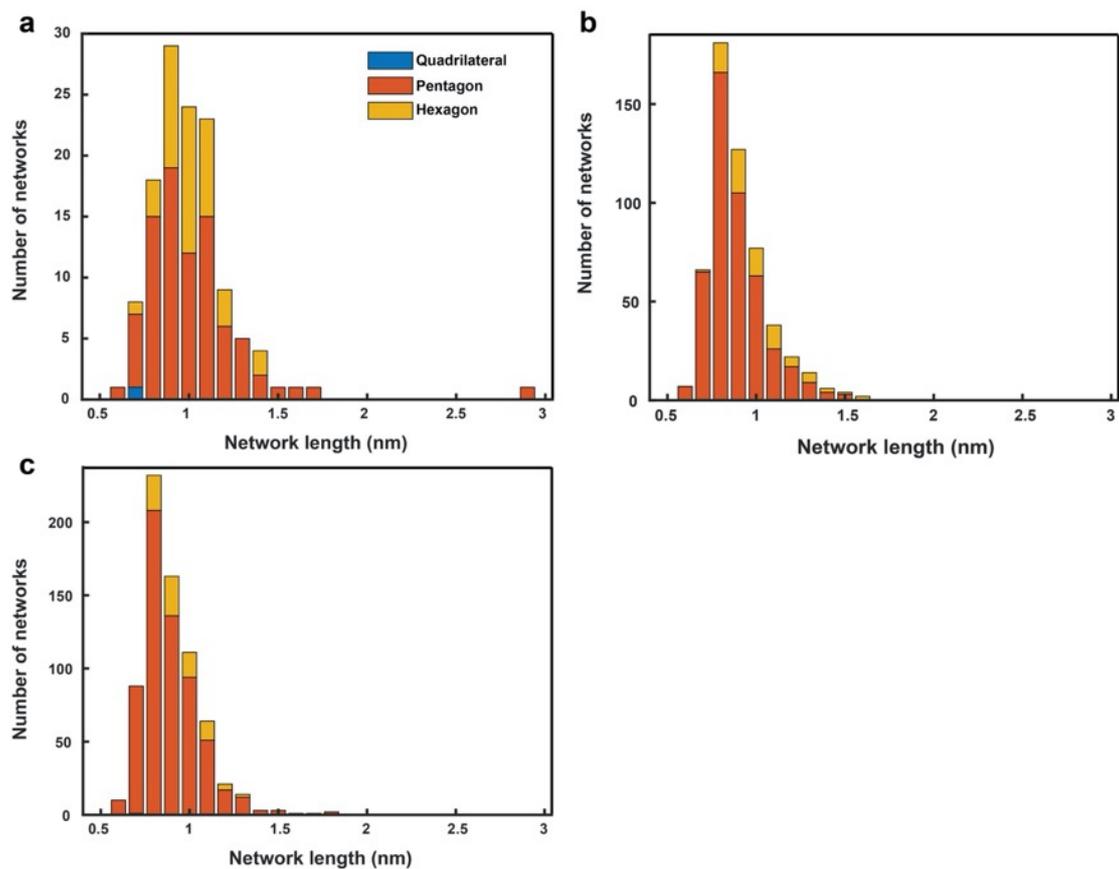

**Supplementary Fig. 15 | Analysis of the quadrilateral, pentagonal, and hexagonal bipyramid network length.** Population of the quadrilateral (blue), pentagonal (orange), and hexagonal bipyramid (yellow) networks as a function of their length in the amorphous Ta thin film (**a**), Pd$_1$ (**b**) and Pd$_2$ (**c**) nanoparticle, where the network length is defined as the distance along the longest direction of each network.

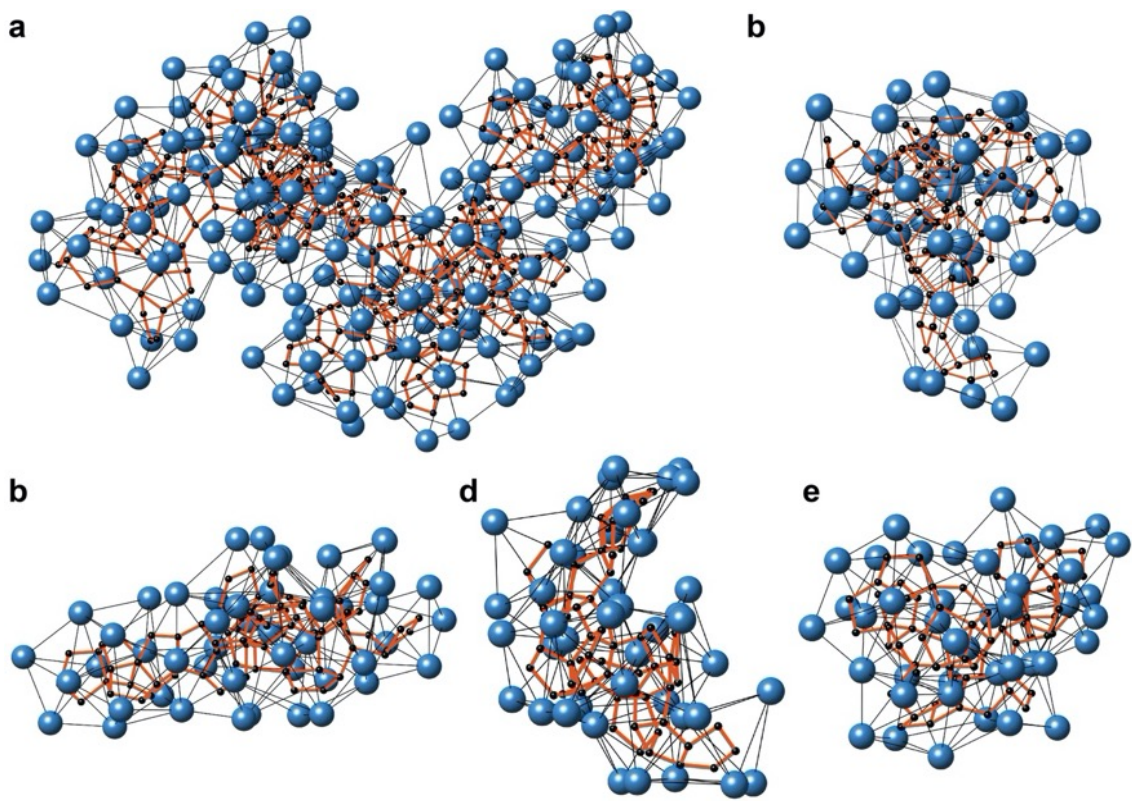

**Supplementary Fig. 16 | Five largest PBNs in the amorphous Ta thin film**, which contain (**a**) 135, (**b**) 35, (**c**) 34, (**d**) 28 and (**e**) 24 pentagonal bipyramids, respectively.

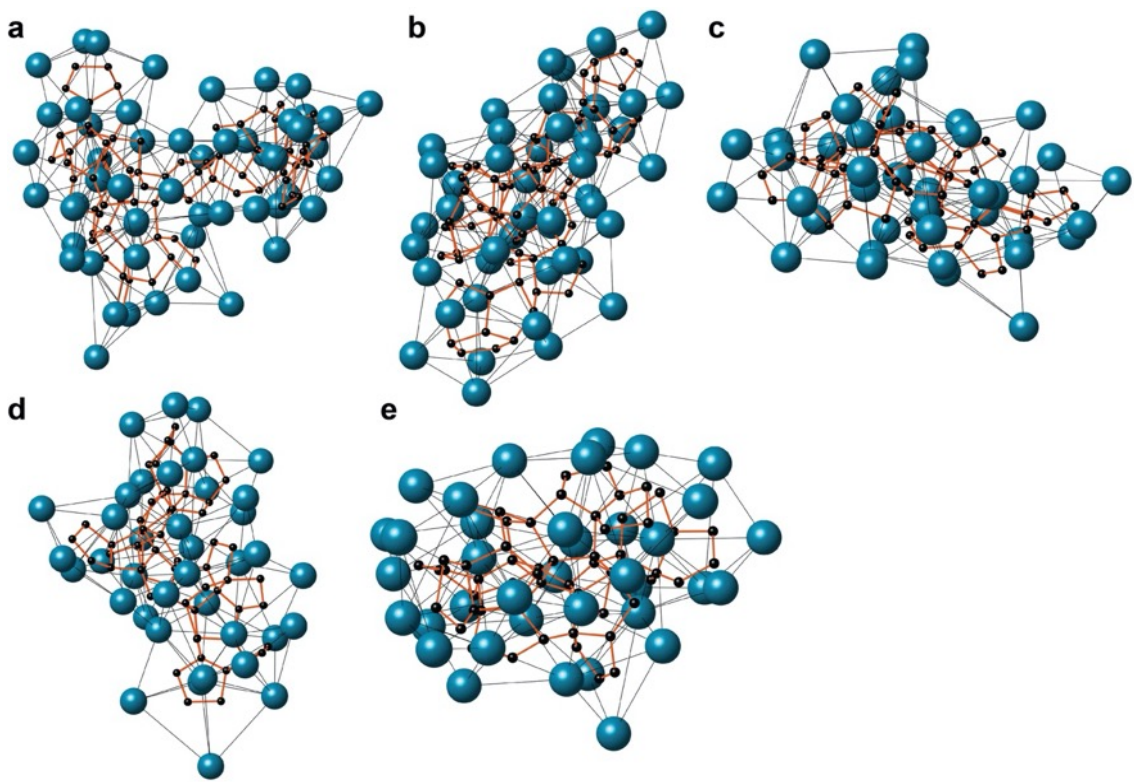

**Supplementary Fig. 17 | Five largest PBNs in the amorphous Pd₁ nanoparticle**, which contain (**a**) 31, (**b**) 28, (**c**) 25, (**d**) 24 and (**e**) 24 pentagonal bipyramids, respectively.

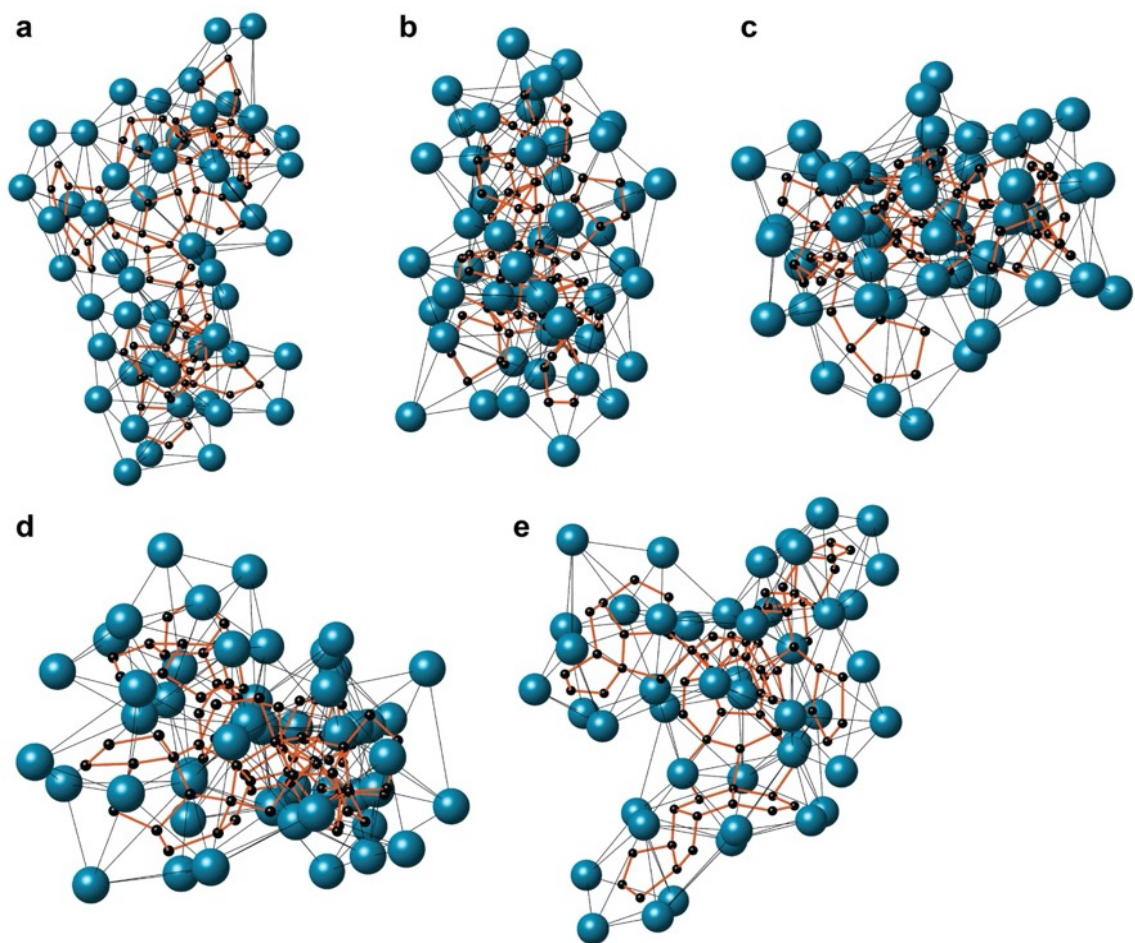

**Supplementary Fig. 18 | Five largest PBNs in the amorphous Pd₂ nanoparticle**, which contain (**a**) 32, (**b**) 31, (**c**) 30, (**d**) 30 and (**e**) 29 pentagonal bipyramids, respectively.

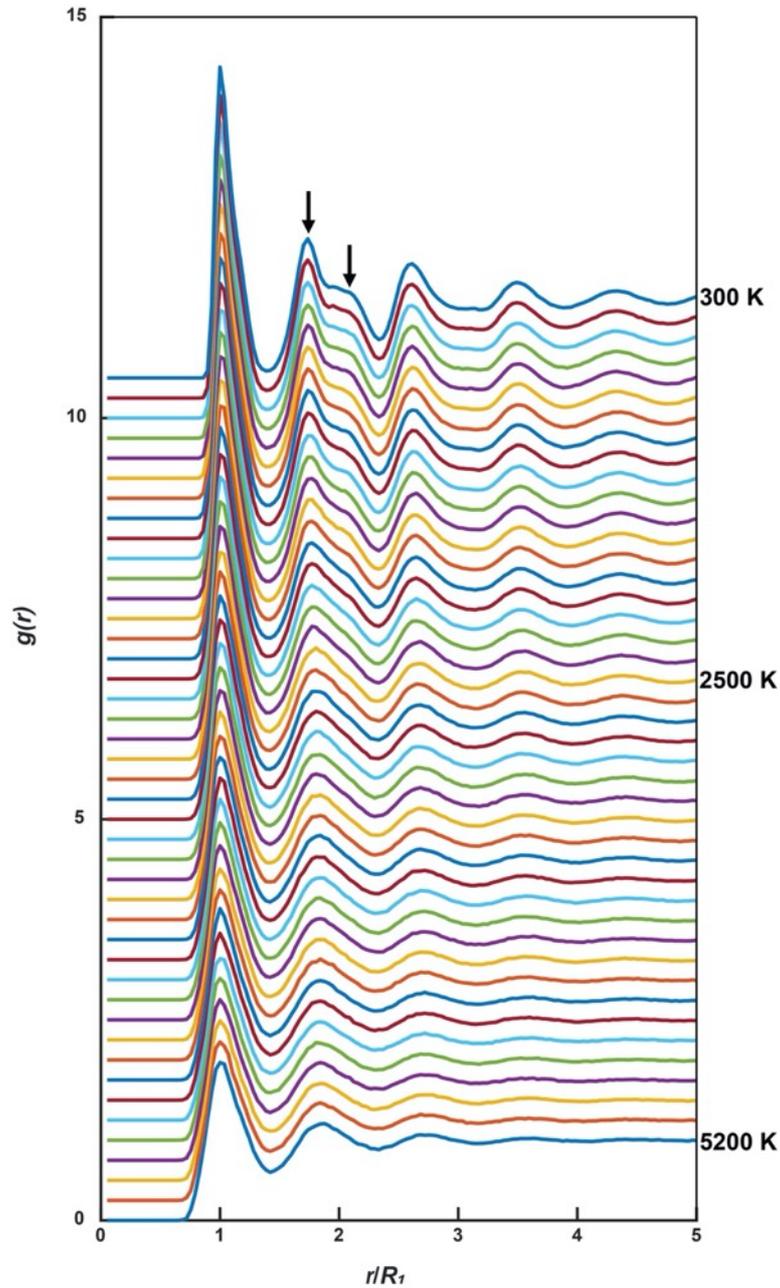

**Supplementary Fig. 19 | PDFs of MD simulated Ta structures during the quench from a liquid (5200 K) to a metallic glass state (300 K).** At 5200 K, the PDF of the Ta liquid resembles those of the three experimental amorphous materials (Fig. 1c). At 300 K, the splitting of the 2nd and 3rd peaks in the PDF indicates the formation of the Ta metallic glass (arrows). From the free volume vs temperature graph, the glass transition temperature ($T_g$) is determined to be 2496 K.

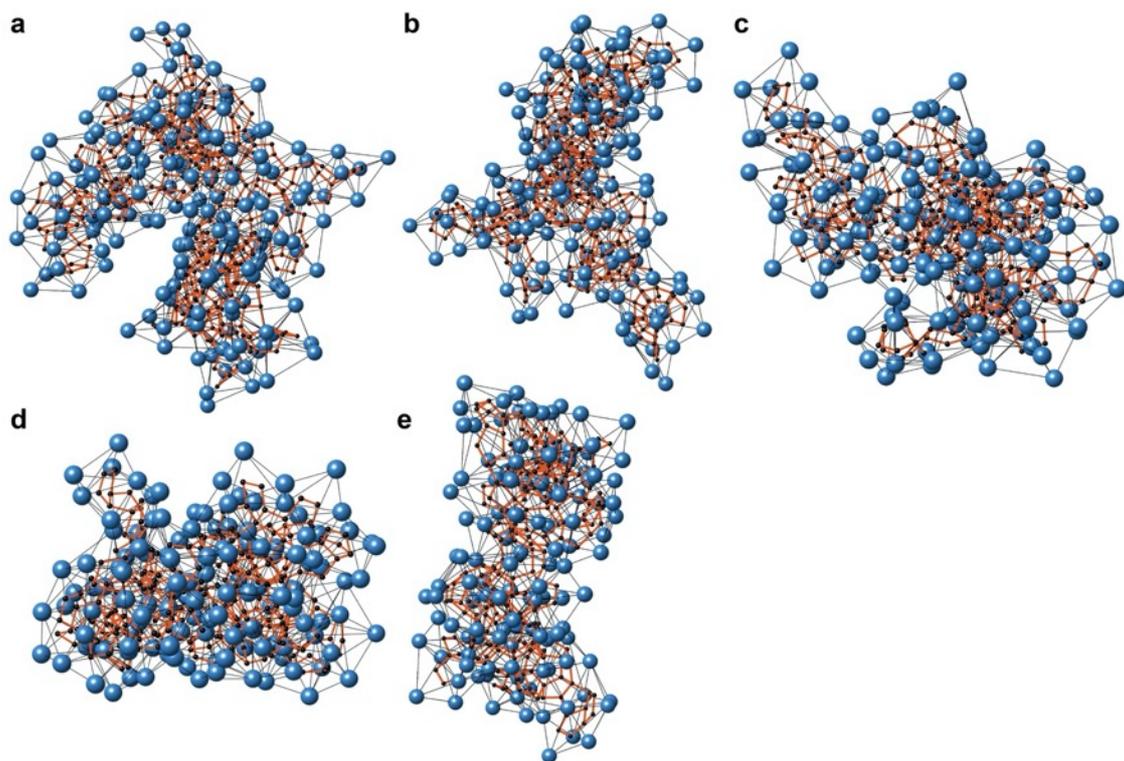

**Supplementary Fig. 20 | Five largest PBNs in the MD simulated Ta liquid at 5200 K**, which contain (**a**) 146, (**b**) 132, (**c**) 120, (**d**) 118 and (**e**) 105 pentagonal bipyramids, respectively.

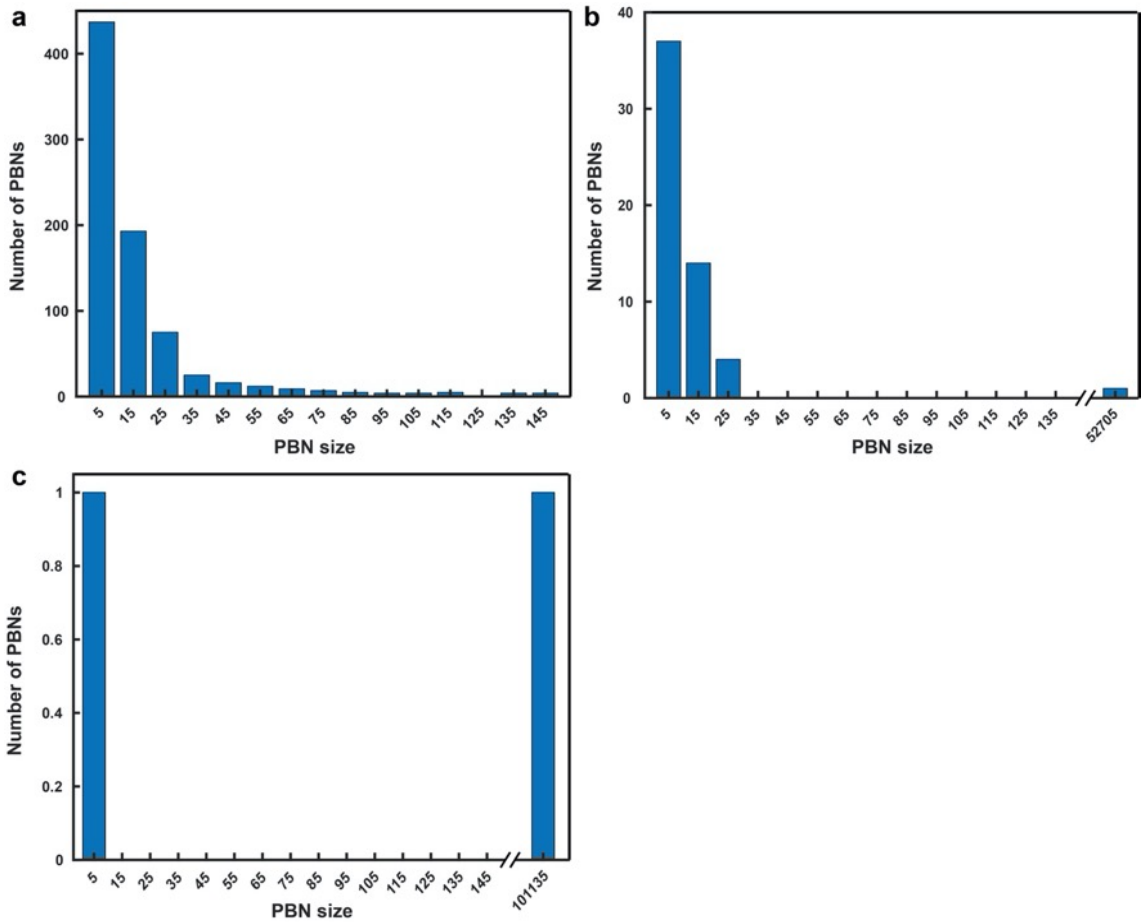

**Supplementary Fig. 21 | Histogram of the PBN size for MD simulated Ta structures at different temperatures.** Number of PBNs as a function of the network size in MD simulated Ta structures at (**a**) 5200K, (**b**) 2500 K and (**c**) 300 K. During the quench from a liquid to metallic glass state, the PBNs rapidly grow in size. At 300 K, a giant PBN is formed that extends to the entire Ta metallic glass.

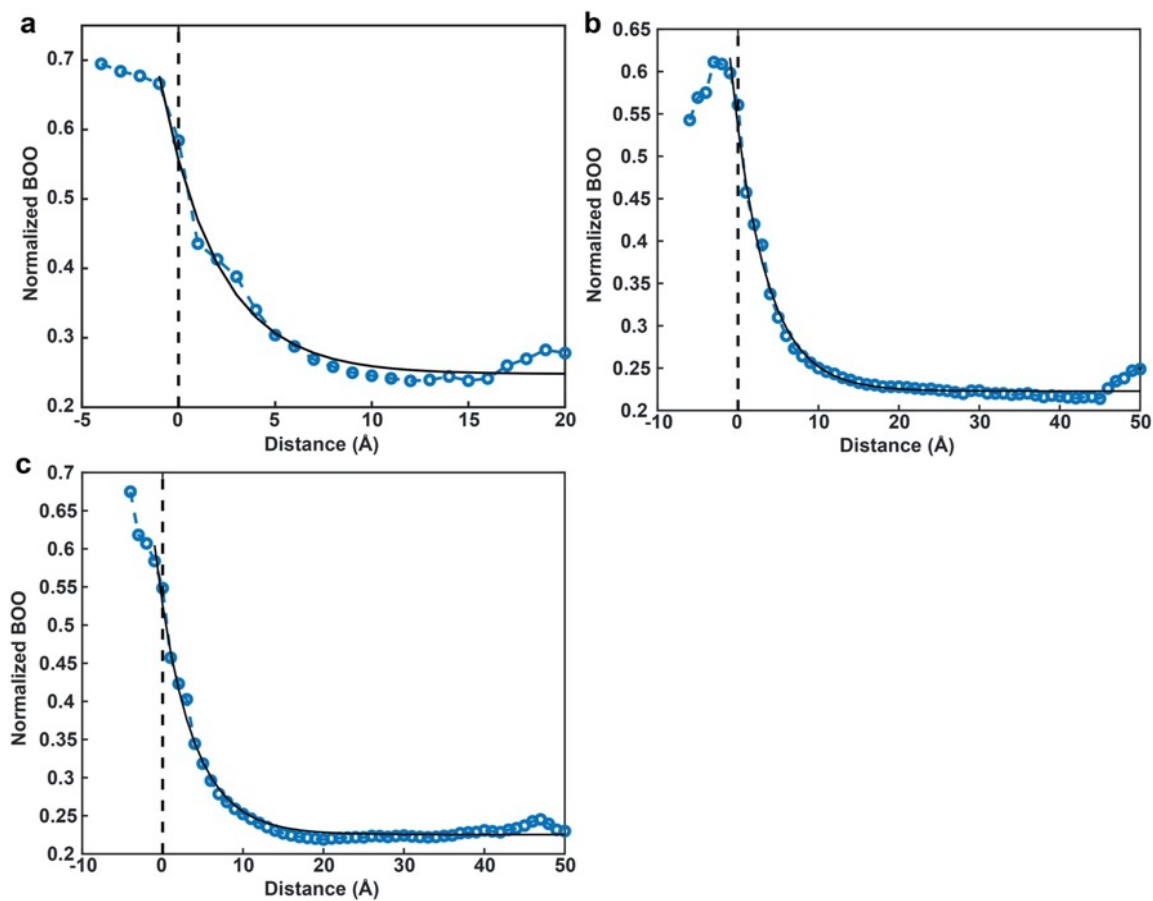

**Supplementary Fig. 22 | Quantifying the crystalline-amorphous interface.** Distribution of the normalized BOO (blue circles) as a function of the perpendicular distance to the surface of the crystal nuclei (dashed line) for the Ta film (**a**), $Pd_1$ (**b**) and $Pd_2$ (**c**) nanoparticle. The experimental data points (blue circles) were fitted with an exponential decay function (solid curves) to determine the characteristic width of the crystalline-amorphous interface to be 3.0, 4.2 and 4.3 Å for the Ta film, $Pd_1$ and $Pd_2$ nanoparticle, respectively (Methods).

## Supplementary Video legends

**Supplementary Video 1**. Experimental 3D atomic model of the amorphous Ta thin film. The reconstructed volume of the thin film consists of 6615 disordered atoms (in blue) with several crystal nuclei of 1669 atoms on the surface (in grey).

**Supplementary Video 2**. Experimental 3D atomic model of the amorphous $Pd_1$ nanoparticle. The nanoparticle consists of 51170 disordered atoms (in green) with several small crystal nuclei of 1138 atoms on the surface (in grey).

**Supplementary Video 3**. Experimental 3D atomic model of the amorphous $Pd_2$ nanoparticle. The nanoparticle consists of 74893 disordered atoms (in green) with several small crystal nuclei of 1345 atoms on the surface (in grey).